\documentclass[sigconf,nonacm,balance=false]{acmart}
\usepackage{popets}
\usepackage{tikz}
\usepackage{amsmath}
\usepackage{filecontents}
\usepackage{booktabs} % For better rule lines
\usepackage{caption} 
\usepackage{makecell}  % Required for '\makecell'
\usepackage{colortbl}
\usepackage{array}
\usepackage{tabularx}
\usepackage{ifthen}
\usepackage{colortbl}
\usepackage{xcolor}
\usepackage{listings}
\usepackage{pifont}
\usepackage{booktabs} % For formal tables
\usepackage{cleveref}
\usepackage{graphicx}
\usepackage{subcaption}
\usepackage{placeins}
\usepackage{stfloats}

\definecolor{delim}{RGB}{20,105,176}
\definecolor{numb}{RGB}{106, 109, 32}
\definecolor{string}{rgb}{0.64,0.08,0.08}

\lstdefinelanguage{json}{
    showspaces=false,
    showtabs=false,
    breaklines=true,
    postbreak=\raisebox{0ex}[0ex][0ex]{\ensuremath{\color{gray}\hookrightarrow\space}},
    breakatwhitespace=true,
    basicstyle=\ttfamily\small,
    upquote=true,
    morestring=[b]",
    stringstyle=\color{string},
    literate=
     *{0}{{{\color{numb}0}}}{1}
      {1}{{{\color{numb}1}}}{1}
      {2}{{{\color{numb}2}}}{1}
      {3}{{{\color{numb}3}}}{1}
      {4}{{{\color{numb}4}}}{1}
      {5}{{{\color{numb}5}}}{1}
      {6}{{{\color{numb}6}}}{1}
      {7}{{{\color{numb}7}}}{1}
      {8}{{{\color{numb}8}}}{1}
      {9}{{{\color{numb}9}}}{1}
      {\{}{{{\color{delim}{\{}}}}{1}
      {\}}{{{\color{delim}{\}}}}}{1}
      {[}{{{\color{delim}{[}}}}{1}
      {]}{{{\color{delim}{]}}}}{1},
}

\newcommand{\name}[0]{\textsc{PriDrive }}

\newboolean{showOEM}
\setboolean{showOEM}{false} % Set to false to hide OEM names
\newcommand{\OEM}[1]{%
  \ifthenelse{\boolean{showOEM}}%
    {%
    \ifthenelse{\equal{#1}{A}}{GM}{%
    \ifthenelse{\equal{#1}{B}}{Honda}{%
    \ifthenelse{\equal{#1}{C}}{Polestar}{%
    \ifthenelse{\equal{#1}{D}}{Volvo}{Unknown}}}}%
    }%
    {OEM #1}%
}

\setcopyright{popets}
\copyrightyear{2025}

\acmYear{2025}
\acmVolume{2025}
\acmNumber{2}
\acmDOI{XXXXXXX.XXXXXXX}
\acmISBN{}
\acmConference{Proceedings on Privacy Enhancing Technologies}
\begin{document}

% \title{I Know What You Did (\textit{In Your Car}) Last Summer: Privacy Implications of Android Automotive OS}
\title{Analyzing Privacy Implications of Data Collection in Android Automotive OS}

%%%%%%%%%%%%%%%% Authors' Info %%%%%%%%%%%%%%%%%

\author{Bulut Gözübüyük}
\email{bgozubu@clemson.edu}
\affiliation{
  \institution{Clemson University}
  \city{Clemson}
  \country{SC, USA}}

\author{Brian Tang}
\email{bjaytang@umich.edu}
\affiliation{
  \institution{University of Michigan}
  \city{Ann Arbor}
  \country{MI, USA}}

\author{Kang G. Shin}
\email{kgshin@umich.edu}
\affiliation{
  \institution{University of Michigan}
  \city{Ann Arbor}
  \country{MI, USA}}

\author{Mert D. Pesé}
\email{mpese@clemson.edu}
\affiliation{
  \institution{Clemson University}
  \city{Clemson}
  \country{SC, USA}}

%oem a -> GM
%oem b -> Honda
%oem c -> Polestar
%oem d -> Volvo

%-------------------------------------------------------------------------------
\begin{abstract}
%-------------------------------------------------------------------------------

% Android Automotive OS (AAOS) has proliferated 
% through most modern vehicles, becoming one of the biggest players 
% in the in-vehicle infotainment (IVI) market, integrated in over 
% 100 million vehicles. \mert{Switch to privacy too abrupt, talk first about data collection.} 

Modern vehicles have become sophisticated computation 
and sensor systems, as evidenced by advanced driver assistance 
systems (ADAS), in-car infotainment, and autonomous driving capabilities. 
They collect and process vast amounts of data through various embedded 
subsystems. One significant player in this landscape is Android 
Automotive OS (AAOS), which has integrated into over 100M vehicles and 
has become a dominant force in the in-vehicle infotainment (IVI) market. 
With this extensive data collection, privacy concerns have become 
increasingly crucial. The volume of data gathered by these systems raises 
questions about how this information is stored, used, and protected, 
making privacy a critical issue for manufacturers and consumers. 
However, very little has been done on vehicle data privacy. 
This paper focuses on the privacy implications of AAOS, 
examining the exact nature and scope of data collection and 
the corresponding privacy policies from the original equipment 
manufacturers (OEMs). We develop a novel automotive privacy analysis tool called \name which employs three methodological approaches: 
network traffic inspection, and both static and dynamic analyses 
of Android images using rooted emulators from various OEMs. 
These methodologies are followed by an assessment of whether 
the collected data types were properly disclosed in OEMs and 
3rd party apps' privacy policies (to identify any discrepancies 
or violations). This allows for a thorough evaluation of OEMs' 
adherence to their stated privacy policies. 
Our evaluation with static and dynamic analyses
on three different OEM platforms reveals that some OEMs collect 
much more data than others. OEM A collects vehicle speed at a
sampling rate of roughly 25 Hz. Meanwhile, other properties such as
model info, climate \& AC, seat data, and others are collected in a
batch 30 seconds into vehicle startup.
In addition, several vehicle property types were collected without 
disclosure in their respective privacy policies. For example,
OEM A's policies only cover 110 vehicle properties or 13.02\% of
the properties found in our static analysis.
Finally, an analysis of the data usage purposes declared in 
privacy policies indicates that data is being used for 
advertising, insurance, financing, and is shared with 3rd parties.
%CCCCC Mention some #s

% Through our detailed static and dynamic analyses using rooted emulators from various OEMs, we assess potential data leakage and privacy vulnerabilities within AAOS.

%[https://www.androidpolice.com/2021/05/19/android-automotive-is-coming-to-a-lot-more-vehicles-this-year-heres-which-brands-are-supporting-it-so-far/]. 

% The paper uses three approaches: Network traffic inspection, static and dynamic analysis of integrated APKs, followed by a privacy policy consistency assessment after these analyses are complete. 

% The urgent need to address privacy flaws arising from the widespread adoption of AAOS motivates our investigation. Our analysis specifically examines the privacy policy violations by various OEMs, identifying the most significant concerns. 

%Not more than 200 words, if possible, and preferably closer to 150.
\end{abstract}

% \keywords{Android, automotive, privacy, infotainment, data collection}
% \keywords{Android Automotive OS, vehicle data privacy, in-vehicle infotainment, OEM compliance, privacy policies, network analysis, static analysis, dynamic analysis, privacy violations, automotive cybersecurity, telemetry data, Man-in-the-Middle, privacy impact assessment.}

\maketitle

%------------------------------------------------------------------------
\section{Introduction}
\label{sec:introduction}
%------------------------------------------------------------------------
%DDDD

Modern vehicles have been increasingly connected and have introduced 
new digital functionality, such as in-vehicle entertainment systems. 
With the development of advanced driver assistance systems (ADAS), 
in-car infotainment and, more recently, autonomous driving 
capabilities, vehicles are no longer just mechanical 
machines --- they have become sophisticated computational 
and sensor systems.

\begin{figure}[t]
    \centering
    \includegraphics[width=\columnwidth]{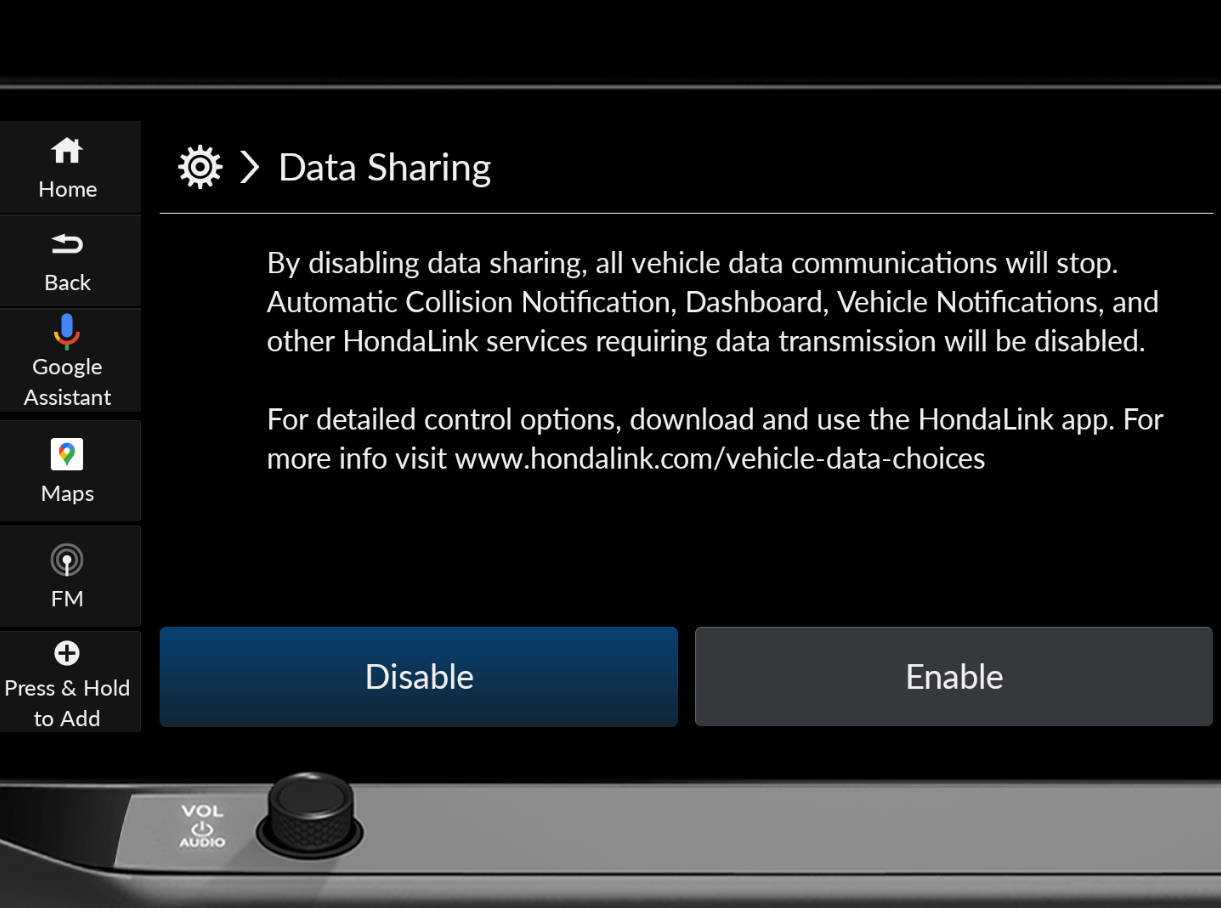}
    \caption{Honda IVI System with Android Automotive OS \cite{honda_emu}}
    \label{fig:teaser}
\end{figure}

Yet, the rich data collected by vehicle sensors poses a significant 
threat to personal privacy. Cars have already been shown to be 
very lucrative sources for data collection. Mozilla's research 
on numerous car OEMs, privacy policies revealed serious privacy 
infringements~\cite{mozilla2023privacy}. Vehicles from brands 
such as Nissan, Volkswagen, Toyota, and more, were shown to 
collect sensitive data such as sexual activity, immigration status, 
race, facial expressions, weight, health and genetic information, 
as well as driving routes. An article by Molla {\em et al.}~\cite{Molla2024May}
explored the opt-out process for a Honda 
vehicle's data collection settings, a process that was full of 
calling dealerships and dark patterns --- UIs designed to be 
challenging to navigate~\cite{bosch2016tales,gray2018dark}. 
Another investigation probed into General Motor's (GM) data 
partnership with OnStar's Smart Driver program, a third-party 
company that collects driving behavior~\cite{Hill2024Apr}. 
Their data-collection program was initiated by dealerships 
during car sales or via rewards programs without clear 
communication with buyers. Despite this, data-collection practices 
led to sharing driver safety data with insurance companies, which led 
to driving up users' premiums. This incident contributed to federal 
lawsuits and compelled GM to cease their 3rd-party data sharing 
practices and a reassessment of their privacy policies~\cite{Hill_2024}. 
Recently, consumers have become increasingly wary of their 
vehicle data privacy. A survey conducted by Baumgartner 
{\em et al.}~\cite{baumgartner2024kaspersky} reported that 71\% 
of drivers considered buying an older vehicle to avoid 
intrusive data collection.
% \mert{One thing I'd add  here: If you buy a vehicle with AAOS, during
% first setup you HAVE to agree to their PPs, otherwise you cannot use the IVI.
% I have seen cases where you cannot opt out at all.}

One particular computing framework has become popular in new 
vehicles. Google's Android Automotive operating system (AAOS) 
has become a well-liked foundation for the upcoming generation 
of connected cars. It has already been adopted by OEMs, such as 
Acura/Honda \cite{hondagoogle}, GM \cite{gmgoogledev}, Ford/Lincoln 
\cite{fordgoogle}, Volvo/Polestar \cite{volvogoogle}, Volkswagen, 
Stellantis \cite{snappauto}, and many more. AAOS offers a full 
range of features, combining entertainment, communication, 
navigation, and vehicle operations into a user-friendly system. 
Although these features offer quality-of-life improvements and 
a highly specialized user experience, they also pose potential 
privacy risks. Due to AAOS's capabilty to gather, process, and 
send large amounts of personal data, it is essential to ensure 
consumers' information gathered from vehicles is not used without 
their consent. In certain AAOS systems that we surveyed, 
users are required to accept the platform's Privacy Policies during 
the initial setup process. Failure to accept these policies will 
result in the inability to utilize the infotainment system.

In this paper, we assess the type and scope of data collection 
and the default privacy settings on both OEM apps and 3rd-party 
apps. We investigate AAOS's underlying behavior, using rooted 
emulators from various OEMs to provide a detailed perspective 
on data leakage and potential privacy vulnerabilities. 
For example, in the permission framework of 
Android Automotive, we investigate situations where a third-party 
app may have access to the vehicle data. Such data, like the RPM 
and gear position, can be used to derive the vehicle speed value, 
bypassing user consent requirements for higher-sensitivity data 
types. As a result, aggregating lower-permission information can 
be used to forecast strict-permission data like vehicle speed or 
driving routes~\cite{pese2020security}. The inferred data, such 
as location from speed or other attributes, might enable location 
tracking and constitute a severe invasion of privacy~\cite{dewri2013inferring, zhou2017speed, pese2020spy}. 

\begin{figure}[t]
    \centering
    \includegraphics[width=\columnwidth]{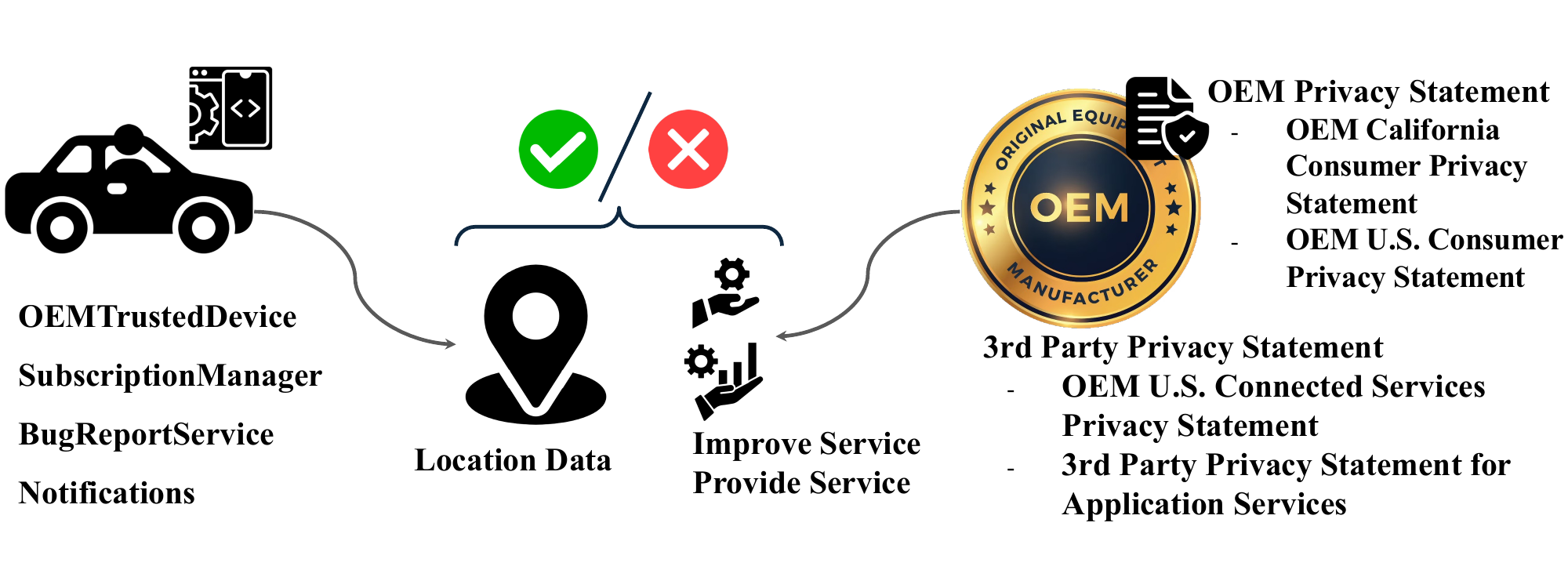}
    \caption{High-level system overview}
    \label{fig:highlevel}
\end{figure}

We investigate the following \textbf{research questions} and make 
contributions that improve upon the state-of-the-art (SOTA) 
in Android Automotive and vehicle privacy:

\begin{itemize}
\item[\textbf{RQ1}:] From permissions and properties in APK source code, 
what insights can we derive about potentially collected data, 
including the permissions and properties
% and URLs/IP addresses 
that are present? It should be noted that the presence of these 
elements may suggest but do not confirm data collection.

\item[\textbf{RQ2}:] Through the use of dynamic analysis techniques via 
Frida, we aim to understand which callback functions to the in-vehicle 
network are active during the runtime execution of an Android 
Automotive application, and what indicators can be collected 
through monitoring. However, the data collected is not exhaustive, 
and its presence does not necessarily mean that data has been 
or will be collected or shared.

\item[\textbf{RQ3}:] How can we determine the data is transferred 
and the contents included in the payload using Man-in-the-Middle (MITM) 
network analysis? We would like to seek clear evidence of data 
transportation.

\item[\textbf{RQ4}:] Can we analyze the data to find patterns and 
correlate data properties found in the static and dynamic analyses 
with those disclosed in privacy policies? By processing the privacy 
policies, we can determine data usage purposes for specific data types.
\end{itemize}

%{\bf Contributions:}
We conduct a unique, extensive investigation into AAOS's 
privacy features. This study makes use of an automated vehicle
privacy analysis tool we created, called \name, enabling a methodical and in-depth 
analysis of privacy issues with AAOS. Our analysis into AAOS
provides the first comprehensive look into OEMs' data privacy practices,
identifying weaknesses and areas for improvement. Our tool serves as 
a vital resource for OEMs, researchers, developers, and 
regulators studying vehicle privacy. The tool 
automates much of the emulation setup and OEM/3rd-party app 
interactions, allowing for plug-and-play testing of 
OEM-specific app stores. The tool can then be used by (1) OEMs to 
ensure 3rd-party apps meet privacy requirements or (2) regulators 
and researchers to evaluate OEMs' data collection practices.

This paper makes the following \textbf{contributions}:
\begin{itemize}
\item A novel toolchain called \name that combines our custom 
  static analysis methodologies, mitmproxy's MITM network traffic 
  interception and HTTPS decryption capability, Logcat's Android 
  logging features, OEM-provided emulators, and Frida's dynamic instrumentation 
  capabilities to perform systematic data flow analysis within AAOS. 
  For instance, we find that OEM A collects many types of vehicle properties with 
  over 6 times as many property occurrences in all but one data category. 
  Vehicle speed and model/make information is also gathered by maps 
  apps of two OEMs. Further details can be found in \cref{subsec:design_emulation}.
\item An evaluation of the consistency between data collected and OEM privacy policies. 
A large language model (LLM) then 
  maps the vehicle properties collected from these emulators to 
  data types disclosed in privacy policies. The tool identifies 
  any discrepancies, omissions, or inconsistencies between the 
  policies and the collected data. We find that the climate and 
  comfort data type is not disclosed in all the analyzed privacy 
  policies, and every OEM uses or shares data for privacy-intrusive 
  applications (insurance, financing, advertising, law enforcement, 
  telematics). Further details can be found in \cref{subsec:design_privacy_policy}
\item An analysis of how three different OEMs have implemented  
  AAOS and highlighting the potential privacy effects of various 
  implementation strategies. These include permissions
  requested by OEM apps, vehicle property accesses, 
  and a network traffic analysis. Users and manufacturers can assess 
  the security measures in use by providing a comparison basis 
  provided by this research.
\item Other than examining manufacturer-specific implementations, we also investigated the privacy implications of third-party applications that could be found in the Google Play Store. Our research includes a detailed analysis of various third-party apps that utilize unique VHAL property IDs. This reveals their access to and potential use of specific privacy-sensitive data. Apps differed significantly in the range of property IDs they could access, with some having access to 128 distinct property IDs. These results may suggest potential privacy risks, as these apps may have extensive access to sensitive information independent of OEMs. Further details can be found in \cref{sec:eval_third_party}.
\end{itemize}
%-------------------------------------------------------------------------------
\section{Background and Related Work}
\label{sec:background}
%-------------------------------------------------------------------------------

% \mert{2.1 reads like generic background and 2.2/2.3 more like related work, so I adjusted the section title accordingly.}
Android Automotive (AAOS) is an integrated base Android platform 
for in-vehicle infotainment (IVI) systems
that supports optional 3rd-party Android apps and 
pre-installed OEM apps~\cite{aosp}. 
AAOS runs entirely within the car, directly interfacing with 
the vehicle's Electronic Control Units (ECUs), unlike Android Auto 
and Apple CarPlay, which mirror apps onto the IVI screen. 
AAOS may include Google services and apps, all of which are part 
of Google Automotive Services (GAS)~\cite{Hofmann_Name_2020}. 

GAS for Android Automotive stands out as the best 
infotainment system in vehicles~\cite{theverge}. 
In a consumer survey, this OS scored the highest in the 
infotainment category compared to conventional OEMs. 
Infotainment rankings for AAOS without GAS, which lacks 
Google's integrated apps and services, are the lowest. 
As more vehicles integrate systems like Android Auto and GAS, 
there are growing concerns about data security and privacy as 
a result of the resulting integration of personal devices with 
onboard systems. The majority of OEMs include GAS in their AAOS 
releases to give outside developers access to the Google Play Store 
and the ability to distribute their apps across various car models. 
OEMs such as BMW and Stellantis offer AAOS without a GAS license to 
provide a distinct brand experience and prevent sharing 
telemetry data with Google. 

% Google, OEMs, and third-party apps are the main stakeholders in AAOS, which also facilitates the collection of driving data. Due to strict international data protection laws, third-party apps on AAOS offer a variety but can raise privacy concerns. 

\subsection{AAOS System Design}
\label{subsec:background_aaos}

Android Open Source Project (AOSP) is a stripped-down version of 
Android is accessible to open-source communities, such as automakers, 
enabling them to download the source code and develop unique IVI 
equipment. Google Assistant, Google Navigation, the Google Play 
Store, and other Google-specific apps are all included in the 
different layers AAOS adds to the open-source AOSP OS. 
Automakers must pay for licenses for these Google-made programs. 
Only with the Car-Manager Library is it possible to establish 
communication between the layers of Android applications towards 
Car Services and Communication Stack~\cite{aospaaos}. 
IVI systems benefit from the same security upgrades, architecture, 
and capabilities as Android, with the help of AAOS, which runs on 
AOSP. AAOS adds special modules to the standard Android system 
to interface with the car. By directly connecting to the in-car 
network, such as the Controller Area Network (CAN) bus, the IVI 
can send and receive data. Both OEMs and 
app developers have access to certain data directly from the IVN due to 
AAOS's integration within the vehicle's network. 

% \mert{The following is not well written. You can look at https://medium.com/@imaginationoverflow/android-automotive-3-vehicle-hal-483aeddbf25b to have a better understanding of how to formulate properties.}

\textbf{Vehicle Hardware Abstraction Layer (VHAL)}
is a service within the Android OS that manages and retrieves 
information about vehicle functionalities. It allows the access 
and modification of specific properties in a car's ECU. Additionally, 
it includes the capability to register a callback function to get 
notified when a property changes. Android apps can utilize these 
functionalities using the CarPropertyManager \cite{carPropMan}.

\textbf{Vehicle Properties}
identify features that OEMs can incorporate into their vehicles. 
Each property represents a simplified function of the vehicle.
These characteristics could include the tire pressure, 
engine RPM, or interior temperature. For each property, the VHAL 
provides metadata that describes the information, such as the data 
representation type and how it can be altered by reading, 
writing, or subscribing.

\textbf{Permission Model on Android.}
In the current setup of Android, permissions are requested from 
the user for compliance, and the Google Play Store requires app 
publishers to link their privacy policies. Each program must 
disclose and request permissions for data collection and 
processing. Permissions are specified when necessary for the 
requisite limited data or operations. When an app is installed, 
specific permissions, referred to as install-time permissions, 
are automatically granted. Others, referred to as runtime 
permissions, call for a program to request while it is 
already running. 

% \begin{itemize}
%     \item \textbf{Install-time permissions:} These permissions allow users limited access to data or allow them to do activities with little or no negative impact on the system or other apps. These permissions, which are normal and signature permissions, are automatically granted when the program is installed.
%     \begin{itemize}
%         \item \textbf{Normal permissions:} They allow access to information and activities outside of the app's sandbox yet with no danger to user privacy or the functionality of other apps.
%         \item \textbf{Signature permissions:} These are only available if the app is signed by the same certificate as the OS or other application that defines the permission. 
%     \end{itemize}
%     \item \textbf{Runtime permissions:} These, also referred to as dangerous permissions, provide an app with more extensive access to prohibited data or enable it to carry out more significant operations. Therefore, the app must ask for them during runtime. 
%     \item \textbf{Special permissions:} The platform and OEMs often specify these permissions to regulate access to primarily significant actions, and they are tied to specific application actions.
% \end{itemize}

%comparison table

\newcommand{\cmark}{\ding{51}}
\newcommand{\xmark}{\ding{55}}

% \begin{table}[htbp]
% \caption{Comparison of android privacy analysis tools?}
% \label{tab:michican_comparison}
% \centering
% \small
% \tabcolsep=0.03cm
% \vspace{-0.3cm}
% \begin{tabular}{l c c c c}
% \toprule
% & \textbf{\begin{tabular}[c]{@{}c@{}}VHAL \\ Support\end{tabular}} & \textbf{\begin{tabular}[c]{@{}c@{}}System-wide \\ analysis\end{tabular}} & \textbf{\begin{tabular}[c]{@{}c@{}}Static \\ Analysis\end{tabular}} & \textbf{\begin{tabular}[c]{@{}c@{}}Dynamic \\Analysis\end{tabular}} \\ \midrule
% \begin{tabular}[c]{@{}l@{}}MobSF \cite{mobsf} \end{tabular} & \xmark & \xmark & \cmark & \cmark \\ \hline
% \begin{tabular}[c]{@{}l@{}}PRIDRIVE \end{tabular} & \cmark & \cmark & \cmark & \cmark \\ \hline
% \bottomrule
% \end{tabular}
% \end{table}

\subsection{AAOS Security and Privacy}
\label{subsec:background_related_work_aaos}

Our investigation into AAOS privacy is well-positioned within 
the present research landscape in the field of Android-based 
systems, which primarily concentrate on privacy and security 
issues. This subsection provides a context for our research 
within this developing topic by reviewing relevant literature.

%DDDDD
Understanding the challenges in this rather new research direction 
has been based on Pesé \textit{et al.}'s work on the security of Android 
Automotive OS (AAOS)~\cite{pese2020security}. The paper highlights 
the shift toward integrating third-party apps and the complexities 
of IVI systems. The integration of these external applications, 
which can access vehicular sensor data, previously restricted to 
only the vehicle network, raises significant privacy concerns. 
The authors conduct a brief study into the weaknesses in AAOS 
security and privacy. A follow-up work expands on 
this~\cite{pese2023first}, investigating and revealing that 78\% 
of AAOS apps do not mention all dangerous permissions in their 
privacy policies, indicating compliance issues. This is 
particularly concerning given the integration of these apps 
with vehicle sensor data, which often includes personally 
identifiable information (PII).

Eriksson {\em et al.}'s research complements these findings by 
examining the safety, security, and privacy aspects of in-vehicle 
apps, particularly those developed for AAOS~\cite{eriksson2019road}. 
While as secure as regular Android apps, in-vehicle apps pose 
risks to road safety and user privacy via access to vehicle 
properties and sensor data, necessitating robust vetting processes 
and security measures. They conclude that static analyses and more 
restrictive vetting processes are required for AAOS apps due to the 
elevated risks. Additionally, Bodei {\em et al.}'s analysis of 
privacy policies from various carmakers show that significant 
user and vehicle data is collected, but many policies lack 
comprehensive information on data management~\cite{bodei2023data}.

% \mert{How is this related to AAOS Security and Privacy?}
% Similar privacy risks in linked entertainment systems are 
% illuminated by research on the tracking ecology of over-the-top 
% (OTT) TV streaming devices, such as Roku TV and Amazon Fire 
% TV~\cite{tv_privacy}. The privacy issues in vehicle IVI platforms 
% are similar to the pervasive tracking and data collection methods, 
% which frequently lack strong security safeguards.

Analytics libraries in mobile apps also introduce new privacy 
concerns due to their design in collecting users' in-app 
activities~\cite{analytics_privacy}. Using static and dynamic 
analyses, their system reveals how these libraries gather user 
data and how many programs unintentionally leak personal data. 
This is especially important for AAOS because of similar potential 
risks may be associated with integrating analytics features 
into automobile apps.

\subsection{Privacy Policy Consistency and Compliance}
\label{subsec:background_related_work_privacy}

Privacy policies are essential yet often complex, making them 
hard for users to understand and difficult to parse automatically. 
Harkous {\em et al.} address this with Polisis, one of the pioneering 
works on automated analysis of privacy policies 
that uses a privacy-centric language model and 
neural-network classifiers~\cite{harkous2018polisis}.

Also, privacy policies are often inconsistent with actual data collection 
practices. For example, a company might not claim to collect biometric 
data in their policy, but implement a face-tagging/tracking feature. 
Bui \textit{et al.} developed PurPliance to detect inconsistencies 
between stated data-usage purposes in privacy policies and actual app 
behaviors~\cite{bui2021consistency}. Their analysis revealed significant 
inconsistencies in 69.66\% of apps, indicating widespread disclosure 
issues. Andow \textit{et al.} introduced PolicyLint, identifying 
contradictions within privacy policies and uncovering misleading 
statements in 14.2\% of analyzed policies~\cite{andow2019policylint}.

Subsequently, Andow \textit{et al.} created PoliCheck, which incorporates 
entity-sensitive analysis to distinguish between first- and 
third-party data flows, finding that up to 42.4\% of apps either 
incorrectly disclose or omit privacy-sensitive data 
flows~\cite{andow2020actions}. Bui \textit{et al.} also further 
developed ExtPrivA~\cite{bui2023detection}, which detects discrepancies 
between browser extensions’ data collection behaviors and their 
privacy disclosures, revealing inconsistencies in 820 Chrome extensions, 
as well as OptOutCheck~\cite{bui2022opt}, which analyzes inconsistencies 
between trackers' opt-out options and their actual data practices, 
finding discrepancies in 11 trackers.

\section{Design}
\label{sec:design}

\begin{figure}[h]
    \centering
    \includegraphics[width=\columnwidth]{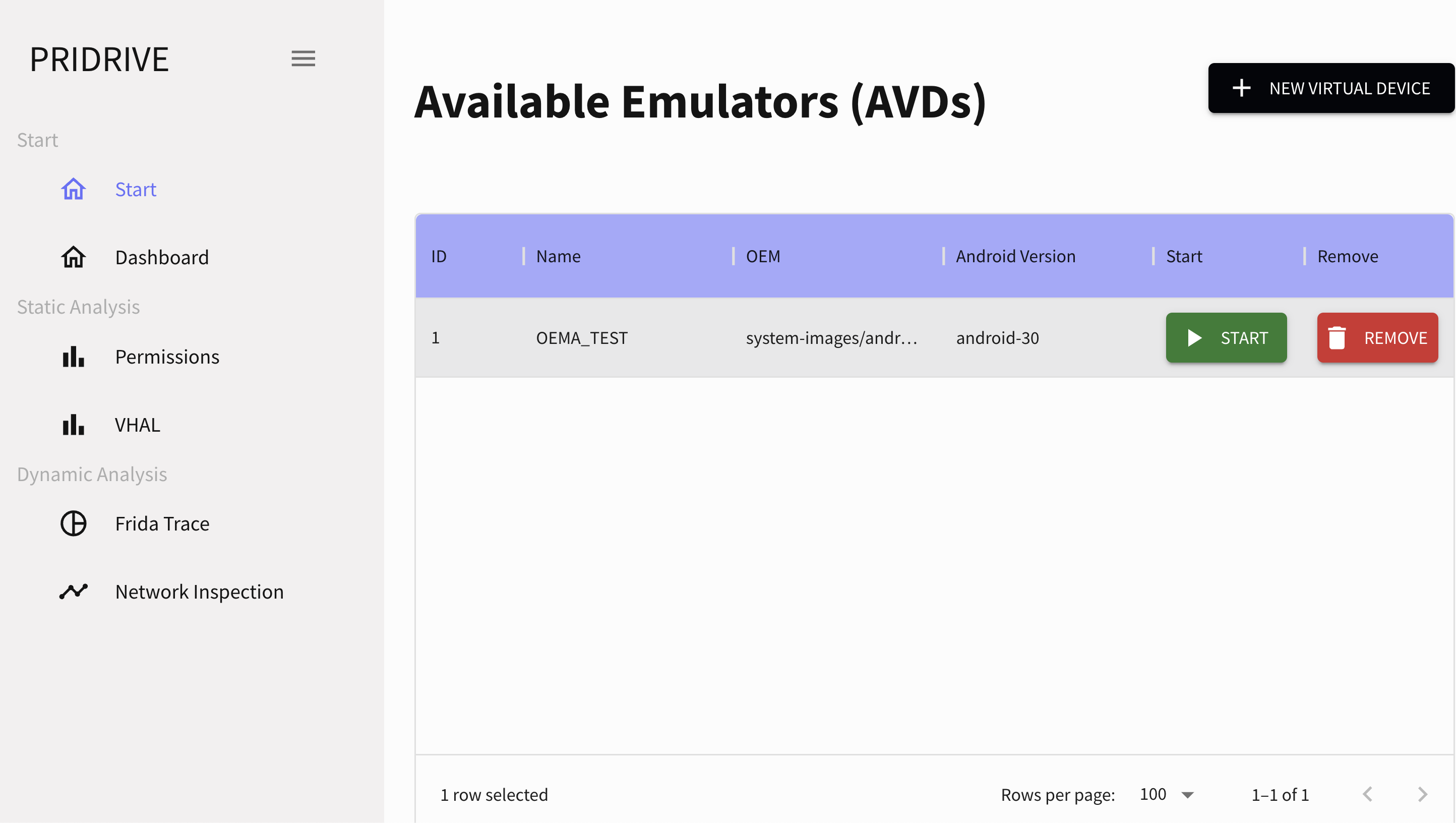}
    \caption{PriDrive Main Page}
    \label{fig:pridrive_home}
\end{figure}

We propose \name, a web application that initiates by setting up an 
emulator and preparing it for subsequent analyses. The system is designed 
to conduct static, dynamic and network analyses. Data gathered from these 
processes are utilized for a privacy policy and permissions evaluation. 
The final output is a comprehensive report that identifies discrepancies 
between the privacy policies and potential privacy/permissions breaches 
detected in the system, which may arise from the network, static, or 
dynamic analysis findings.

\begin{figure}[h]
    \centering
    \includegraphics[width=\columnwidth]{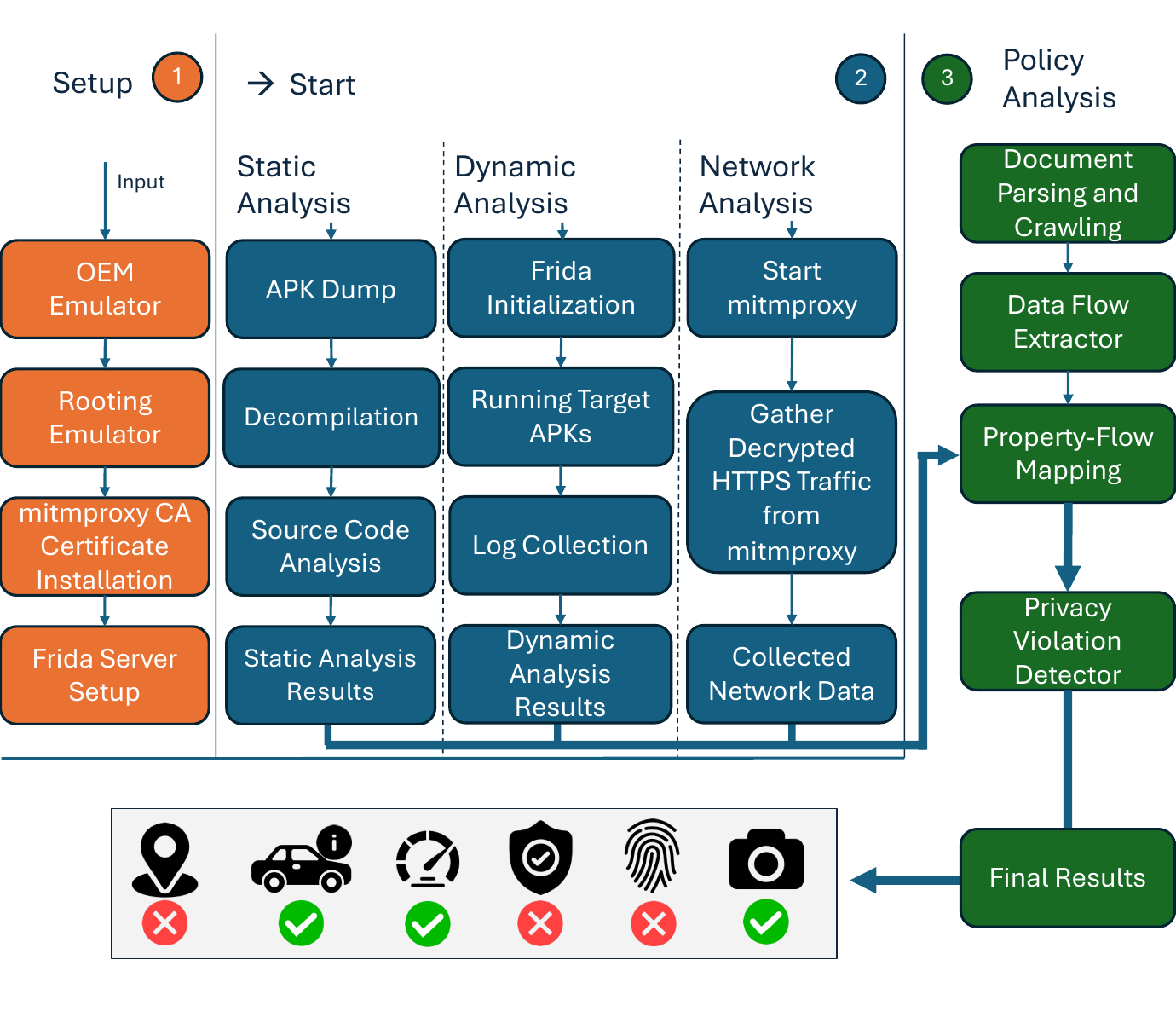}
    \caption{System Design}
    \label{fig:system_overview}
\end{figure}

The proposed tool is designed to work with the Android Software 
Development Kit (SDK) and requires installing command-line tools as a 
prerequisite~\cite{cmdline_tools}. The system architecture employs a 
client-server model, using React JS~\cite{reactjs} with Material-UI 
v5~\cite{mui} for the front end to ensure a responsive and aesthetically 
pleasing user interface. On the back end, it leverages Flask~\cite{flask}, 
a Python-based framework, to handle server-side logic and data management 
efficiently.

\subsection{Emulation, Logging, and Network Components}
\label{subsec:design_emulation}

\name relies on four components to track and analyze data accesses and 
data collection activities: Logcat (logging utility), Frida (app 
debugging/tracing), mitmproxy (decrypt and listens to HTTPS traffic),
and Wireshark (network protocol analyzer).

\textbf{Logcat}~\cite{logcat} gathers system debug data and developer 
log outputs, giving our tool insights into system events, application life 
cycle events, user interactions, warnings, failures, and other pertinent 
information. In the context of our research, Logcat logs are used to track 
particular interactions that may have privacy concerns, performing keyword 
and vehicle property searches that provide information on how data is 
accessed and used by AAOS apps.

The \textbf{Frida} dynamic instrumentation toolset~\cite{Frida}, allows 
real-time viewing and altering of applications' behaviors. 
\name uses it to trace function calls and insert new scripts into the live 
execution environment. In our investigation of AAOS, we used Frida to 
track down and examine function calls and the parameters that accompanied 
them, filtering specifically for data access and utilization.

We examined network interactions in AAOS using a potent tool called 
\textbf{mitmproxy}, among other choices~\cite{mitmproxy}. Mitmproxy acts 
as a man-in-the-middle (MitM) between the application and the server, 
enabling it to decrypt and examine HTTPS, generally encrypted 
communication. We worked on a rooted emulator to successfully intercept 
HTTPS traffic on AAOS and added the mitmproxy certificate to the system's 
trust store. This configuration made it easier to decode otherwise 
encrypted connections, allowing us to examine the information contained 
in network requests and responses.

\textbf{Wireshark} is a well-known network protocol analyzer that we used 
extensively during our network analysis~\cite{wireshark}. A thorough 
picture of network traffic is provided by Wireshark, which allows for 
detailed inspection of hundreds of protocols. From a network interface, it 
provides live packet data, displaying comprehensive details about each 
packet, such as content, protocol, source and destination addresses, and 
more. The network packets sent to and from AAOS were captured and examined 
for our study using Wireshark. As a result, we were able to examine the 
contents of these packets for security research and comprehend the 
underlying network protocols in use. Identification of possible 
vulnerabilities and comprehension of AAOS network behavior were made 
possible by Wireshark's capacity to filter and analyze network traffic.
We investigated the details of transferred data, the associated server 
endpoints, and the frequency and timing of such communications using 
mitmproxy.

\section{Data Processing}
\label{sec:methodology}
%-----------------------------------------------------------------

\subsection{Emulator Setup and System Dump}
% Frida manual interactions with apps and clicking and stuff
\begin{figure}[h]
    \centering
    \includegraphics[width=\columnwidth]{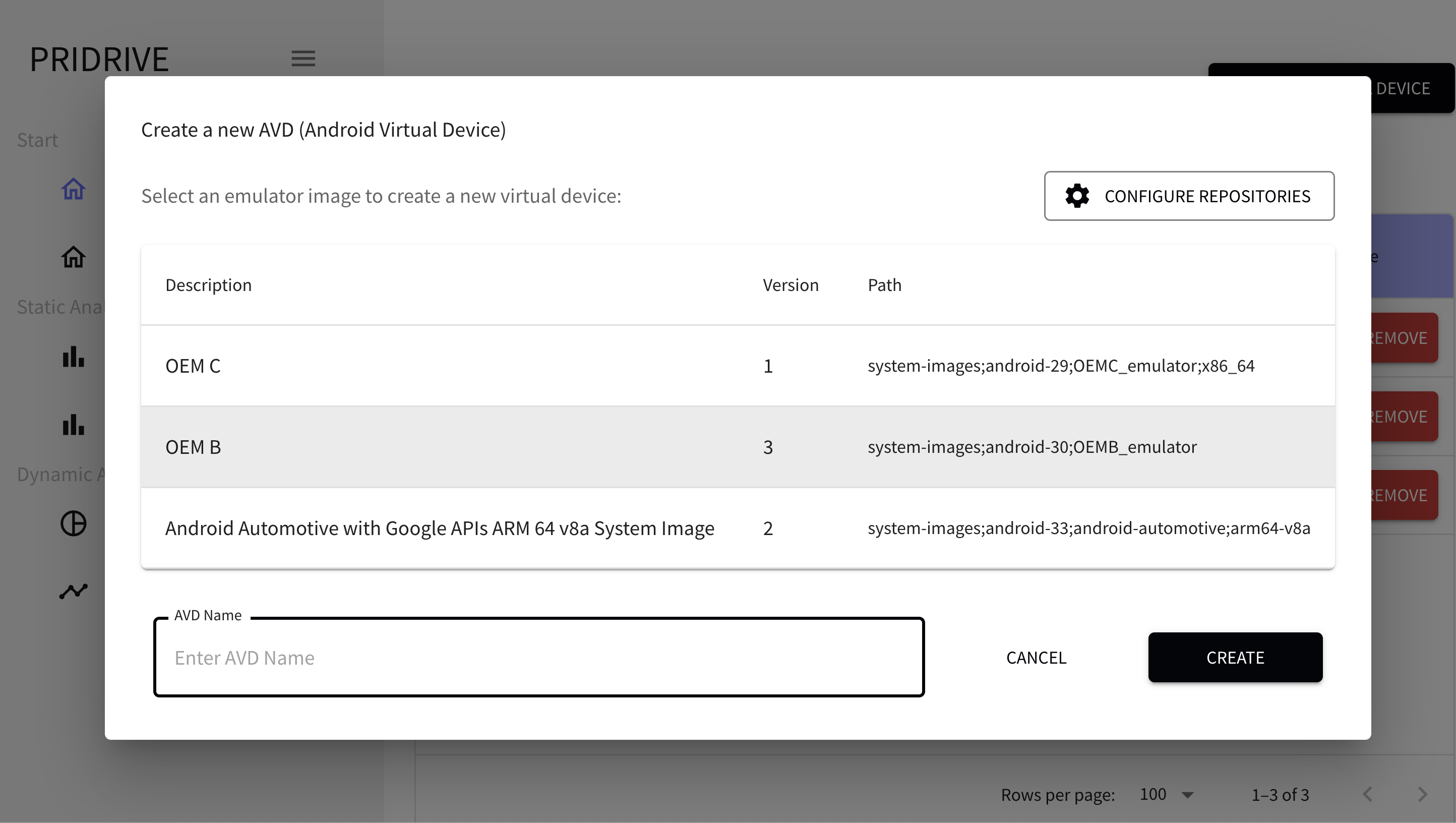}
    \caption{PriDrive emulator (AVD) initialization step}
    \label{fig:pridrive_create_avd}
\end{figure}
    The setup process begins by creating an Android emulator using a 
    publicly available production image from an OEM within Android Studio. 
    Then, we use a \textit{Magisk} module to gain root access and superuser 
    privileges on the emulator. To intercept and analyze network traffic, 
    we reconfigure the root Certificate Authority (CA) certificate of 
    \textit{mitmproxy} to match Android's CA certificate format. We push 
    the modified \textit{mitmproxy} CA certificate to the emulator, 
    allowing \textit{mitmproxy} to intercept all network communication. 
    Lastly, we push the \textit{Frida} server binary to the emulator and 
    grant it executable permissions to allow for instrumentation on 
    application activity. Then, we create a system dump of all 
    applications (APK files) within the emulator. 
    %After the emulator setup, with our Automator script that collects all APK files within the emulator, we created a system dump from that for static analysis.

    We have employed \textit{Frida-Trace} as a dynamic instrumentation tool 
    to log system VHAL calls associated with APK files. We have developed a 
    bash script that starts the emulator, runs the specific app, and starts 
    the Frida tracing for 5 minutes. This instrumentation allowed us to 
    monitor the real-time interaction between the VHAL and the software 
    layer. As a result, we are able to have insights about the behavior of 
    the package files within the operating system. Furthermore, we have 
    embedded two more different data collection methods into our script.
    One collects the \textit{logcat} information, and the other is tcpdump.

    The study procedure is divided into three phases, each of which 
    concentrates on a certain aspect: static analysis, dynamic analysis, 
    and network inspection. The next sections provide a full explanation 
    of these phases in terms of data collecting in this section, while 
    data processing is discussed in the following section.

% \mert{This is a good place to give an overview of the three-phase design. Name the phases. You kinda do that in Figure 3/section 3. There is quite a bit of overlap between Sections 3 and 5. Maybe we can have one section for each step of the pipeline as shown in Figure 3. Sections 3.1, 3.2 and 3.3 can be merged into each respective section and Section 3 itself would just provide a high-level overview of the entire system?}

\subsection{Phase 1A: Static Analysis}
\label{subsec:method_static_analysis}

With the APK dump from each of the emulators, we have developed a pipeline 
that automatically decompiles each APK file and searches for keywords 
within that decompiled source code using jadx \cite{jadx}.

% GPT-4 from OpenAI was used to help create the list of privacy-related keywords used in this section of static analysis \cite{GPT4_2023}. 

% "Could you generate a comprehensive list of keywords focused on privacy, data storage, data transmission, user activity, and vehicle functionalities within the Android Automotive OS ecosystem? I need keywords that touch upon location services, shared preferences, SQL databases, file operations, content resolution, HTTP protocols, web sockets, client-server interaction, device identifiers, cryptography, logging, analytics, geolocation, and permissions. I need them to paste into python3 script to parse decompiled APK file content."

% A multi-step methodology was used in an iterative approach to filter the keywords for this analysis.

% \begin{itemize}
%     \item Initial research: We evaluated AAOS's documentation and current literature to create a list of relevant keywords.
%     \item Source code analysis: We have analyzed a relevant section of the AAOS codebase and decompiled OEM APK source code to find repeating keywords for handling private or sensitive data.
%     \item Industry review: We studied industry standards and whitepapers regarding AAOS and privacy issues.
% \end{itemize}

% \mert{This paragraph is a bit abrupt and unrelated to the flow. Move it somewhere else?}
% Furthermore, developers use emulator detection to control runtime behavior to protect their apps from being analyzed dynamically with emulators, and these techniques are commonly used by malware \cite{emulatordetection}.

In order to assess the functionality available to OEMs, our study parsed 
VHAL property IDs from the system dump. These attributes are defined by 
VHAL and these should be taken into consideration for evaluating potential 
privacy threats.

% \mert{General comment: Do not talk too much about "preparing" or "creating" scripts, but more about what they do. You can mention above once that all steps are automatized. Yes, we are talking about implementation, but this is not a tech report.}
% After a manual inspection of property IDs within the extracted source code, we have identified their presence in specific source files across various vendors in specific APK files. 
We compiled a list of custom property IDs and their corresponding 
descriptions for each OEM as key-value pairs. A sample JSON output of 
VHAL property accesses is illustrated below.

\begin{lstlisting}[language=json, caption=Sample JSON output representing VHAL property ID occurences in specific APK files, basicstyle=\ttfamily\footnotesize]
"ApplicationPackage1Name": {
    "VHAL_PROPERTY1_ID": {
        "description": "VHAL_PROPERTY1_DESCRIPTION",
        "occurrences": PROPERTY1_OCCURENCE_COUNT
    },
    "VHAL_PROPERTY2_ID": {
        ...
},
"ApplicationPackage2Name": {
    ...
\end{lstlisting}

After scrutinizing each APK, as well as the corresponding Android Manifest 
and source code, we examined the custom permission usage of
various OEMs. It is worth noting that system applications are not obligated 
to declare their permissions within the Android Manifest. To address this 
shortcoming, our analysis included the source code of these system 
applications. Subsequently, we assembled a comprehensive list of each OEM 
and each existing APK file, including all the custom permissions that are 
utilized in their respective applications.
% \mert{No reference to listing below.}

\begin{lstlisting}[language=json, caption=Sample JSON output representing vendor permission occurences in specific APK files, basicstyle=\ttfamily\footnotesize]
"ApplicationPackage1Name": {
    "USED_VENDOR_PERMISSION1_DEFINITION",
    "USED_VENDOR_PERMISSION2_DEFINITION",
    ...
},
"ApplicationPackage2Name": {
    ...
\end{lstlisting}

For each OEM, we categorized custom VHAL properties within AAOS into 
six groups for a focused analysis of privacy policies: 

%\section*{Categories}
\begin{itemize}
    \item[\textcolor{red}{A:}] User Preferences and Notifications Settings
    \item[\textcolor{orange}{B:}] Driving Assistance and Mode Security
    \item[\textcolor{yellow}{C:}] Energy and Maintenance
    \item[\textcolor{green}{D:}] Lighting
    \item[\textcolor{blue}{E:}] Diagnostic and Monitoring
    \item[\textcolor{purple}{F:}] Climate and Comfort
\end{itemize}

Each category reflects a distinct functional area in modern vehicles, 
allowing for targeted analysis of privacy policies relevant to each area 
and ensuring comprehensive coverage of all aspects of vehicle operation 
and user interaction. This categorization simplifies the analysis of 
privacy policies by making them more understandable. See 
\Cref{subsec:appdx_data_categories} for more details.

% \mert{I am curious what happened to OEM D.}

% \begin{table}[]
% \begin{tabular}{lllll}
%   & GM & Honda & Volvo & Polestar \\
% A & 1600  & 62    & 22     & 8         \\
% B & 1698  & 20    & 1      & 11         \\
% C & 1076  & 9     & 1      & 0         \\
% D & 72    & 45    & 0      & 0        \\
% E & 1107  & 74    & 0      & 0        \\
% F & 780   & 44    & 16     & 16      
% \end{tabular}
% \end{table}

\begin{table}[h]
\centering
\begin{tabular}{|l|l|l|l|l|}
\hline
\rowcolor[gray]{0.8}
  & \textbf{\OEM{A}} & \textbf{\OEM{B}} & \textbf{\OEM{C}}  \\ \hline
\cellcolor{red!30}[A] User Preferences & 182  & 103     & 39\\ \hline
\cellcolor{orange!30}[B] Driving assistance & 196  & 69      & 47      \\ \hline
\cellcolor{yellow!30}[C] Energy and Maintenance & 136  & 10     & 11        \\ \hline
\cellcolor{green!30}[D] Lighting & 9    & 43    & 29         \\ \hline
\cellcolor{blue!30}[E] Diagnostic and Monitoring & 136  & 158    & 38          \\ \hline
\cellcolor{purple!30}[F] Climate and Comfort & 94   & 52    & 28         \\ \hline
\cellcolor{white}Total Property Count & 753   & 435    & 192         \\ \hline
\cellcolor{white}Total Permission Count & 225   & 338    & 34         \\ \hline
\end{tabular}
\caption{Total Permission and VHAL Property counts in an OEM
emulator by category}
\label{your-label}
\end{table}

\subsection{Phase 1B: Dynamic Analysis}
\label{subsec:method_dynamic_analysis}

\textbf{Classifying \textit{frida-trace} outputs.}
%DDDDDD
After conducting a manual review of VHAL calls in the extracted Frida trace, we determined that these calls can be present in five scenarios. Note that 'PROPERTY\_ID' serves as a more descriptive placeholder for the property ID as it appears in the trace:
\begin{itemize}
\item CarPropertyValue.getPropertyId() <= 'PROPERTY\_ID'
\item ICarProperty$Stub$Proxy.getProperty('PROPERTY\_ID')
\item CarPropertyManagerEx.getProperty('PROPERTY\_ID')
\item .prop = 'PROPERTY\_ID'
\item registerListener: propId is not in config list: 'PROPERTY\_ID'
\end{itemize}

We executed a detailed and comprehensive dynamic analysis by utilizing \textit{frida-trace} outputs. During initial testing, we observed that significant and diverse actions of the applications were recorded within the first few minutes of engagement. After 2-3 minutes, the behavior tended to become repetitive, and we did not observe additional unique property accesses. Thus, data is recorded over periods of five minutes for each application package in JSON format to be processed in section \ref{subsec:eval_dynamic_analysis}. 

\begin{table}[h]
\centering
\begin{tabular}{|l|l|l|l|l|}
\hline
\rowcolor[gray]{0.99}
  & \textbf{\OEM{A}} & \textbf{\OEM{B}} & \textbf{\OEM{C}}  \\ \hline
\cellcolor{red!15}[A] User Preferences & 1600  & 62      & 8\\ \hline
\cellcolor{orange!15}[B] Driving assistance & 1698  & 20      & 11      \\ \hline
\cellcolor{yellow!15}[C] Energy and Maintenance & 1076  & 9     & 0        \\ \hline
\cellcolor{green!15}[D] Lighting & 72    & 45    & 0         \\ \hline
\cellcolor{blue!15}[E] Diagnostic and Monitoring & 1107  & 74    & 0          \\ \hline
\cellcolor{purple!15}[F] Climate and Comfort & 780   & 44    & 16         \\ \hline
\end{tabular}
\caption{Total Custom VHAL Property \underline{occurrences} in Dynamic Analysis of an OEM
emulator by category within 5 minutes}
\label{total_occr_by_oem_prop}
\end{table}

\begin{figure}[ht]
    \centering
    \includegraphics[width=\columnwidth]{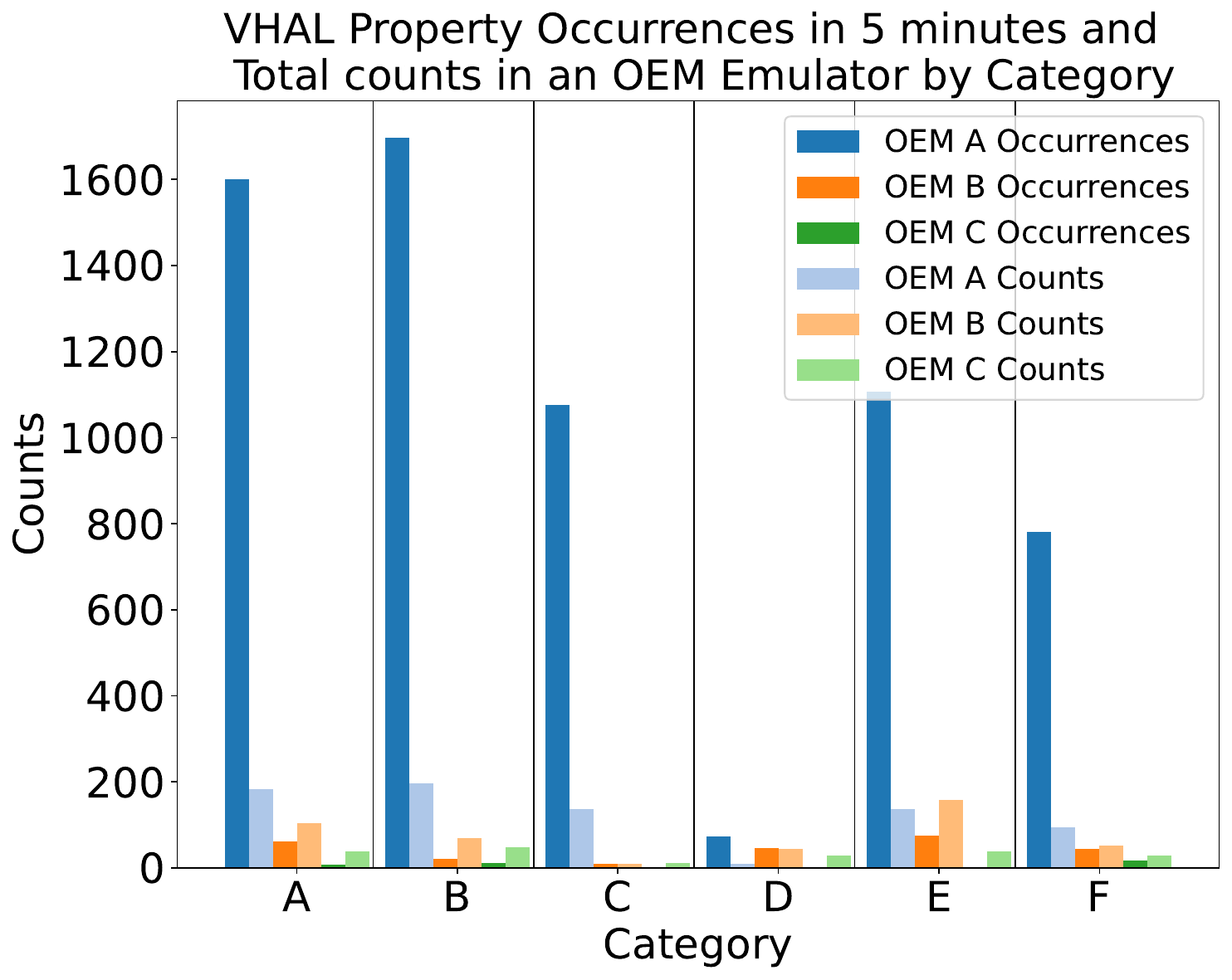}
    \caption{Property Occurrences in 5 minutes and Total counts in an OEM Emulator by Category for Comparison}
    \label{fig:count_occr_prop}
\end{figure}

% \subsubsection{Logcat Log Analysis}

% \subsubsection{Data Analysis and Visualization}

\subsection{Phase 1C: Network Traffic Collection}
%\textbf{Processing Network Packet Capture (PCAP) Files.}
An important component of our investigation was the processing and analysis of the decrypted HTTPS traffic and PCAP files. This section explores the multi-step procedure we used to handle and analyze this network-related data.

\textbf{Parsing and Extraction.}
First, the PCAP files were parsed to retrieve an ordered series of network packets, ensuring that the chronology of the data flow was preserved. The fundamental portion of the data from which we gleaned insights was created by the appropriate headers of each packet, which were examined as part of the parsing process.

Each packet header was processed to separate and identify distinct characteristics such as IP addresses, source and destination, used ports, and underlying protocols. Through recording the network activity, we were able to notice trends in AAOS app behavior. For example, OEM A's V2X service, Radio, and Analytics app were sending the largest amounts of data. Performing reverse IP lookups revealed that the majority of data are collected and sent over network traffic to Google data centers.

The unencrypted HTTPS flows performed a similar dissection in parallel with this process. The hostnames, URLs, HTTP methods (such as GET, POST, DELETE, and PUT), status codes, headers, and payloads of these HTTPS requests and responses were all carefully analyzed. This step is intended to understand the data transfers and requests between the servers and the AAOS, including the types of data being communicated and the frequency of requests.

\textbf{Network Traffic, Process ID, APK Correlation.}
To identify the source of the network traffic, we correlated the network data parsed with the corresponding process IDs (PIDs) captured with each APK file, allowing us to correlate network requests with specific applications. In parallel, the active PIDs and corresponding APKs were mapped simultaneously using \textit{ps} ("process status"). Our tool links the logs gathered from the \textit{netstat} and \textit{ps} commands, the network traffic information from PCAP files, and the decrypted HTTPS flows. \textit{mitmproxy} was also used to collect data such as hostnames, URLs, HTTP methods, status codes, headers, and the body contents of both requests and responses. 

\subsection{Privacy Policy Crawling and Parsing}
\label{subsec:design_privacy_policy}

In order to ensure vehicles are complying with their companies' privacy policies, our toolkit includes an HTML and PDF parser that extracts sentences from privacy policies, allowing fine-grained privacy policy representation and analysis. After separating the policies into sentences, the parser uses an LLM to group sentences into paragraphs and sections to retain context. Our system preserves relevant information within sections and across sentences compared to prior work and traditional privacy policy parsers. After chunking, an LLM extracts data flow representations from each sentence in a privacy policy to identify key attributes relevant for data flow analysis:

\begin{itemize}
    \item Action Verb: Actions taken (e.g., collect, share, use).
    \item Purpose Category: General intent (e.g., marketing, advertising, improving service).
    \item Specific Purpose: Precise reason for the action (e.g., to provide a loyalty program for users to earn and redeem points).
    \item Entity Type: Involved entities (e.g., OEM, third party).
    \item Data Type (Dynamic): Mentioned data types (e.g., personal info, vehicle information, vehicle safety data).
    \item Data Source (Dynamic): Origins of the data (e.g., from a connected mobile device).
    \item Third Party: Involvement of third parties (e.g., sharing or selling data with 3rd party partners).
    \item Exclusion: Any exclusions or limitations on data collection (e.g., except when you opt out of receiving targeted advertisements).
\end{itemize}

Our toolkit maps specific vehicle properties to dynamic data types from privacy policies, allowing vehicle data and sensor data to be mapped to data types disclosed in privacy policies. The vehicle properties and Android permissions collected from each of the \name's components are mapped to a set of data type categories. For the automated parsing and analysis, we manually crawled through OEM privacy policies to collect URLs and PDFs of privacy policies for our tool to scrape. Our system also crawls the Google Play Store to find each 3rd-party and OEM app that is enabled on AAOS and collects their respective privacy policies, enabling an analysis of both OEM and 3rd-party data practices in the vehicle ecosystem.

%-----------------------------------------------------------------
\section{Evaluation of OEM Apps}
\label{sec:eval}

This section investigates the potential privacy concerns 
resulting from the system's communication patterns, APK 
behaviors, and claimed privacy rules. Our investigation was 
based on three separate but related techniques: static analysis, 
dynamic analysis and network traffic monitoring, followed by 
an evaluation of the consistency of privacy policies. 
% \mert{Not quite if I look at the following subsections (where is the static?) and definitely not in that order.}
% \Cref{subsec:eval_static_analysis}

\subsection{Phase 2A: Static Analysis}
\label{subsec:eval_static_analysis}

\begin{figure}[h]
    \centering
    \includegraphics[width=\columnwidth]{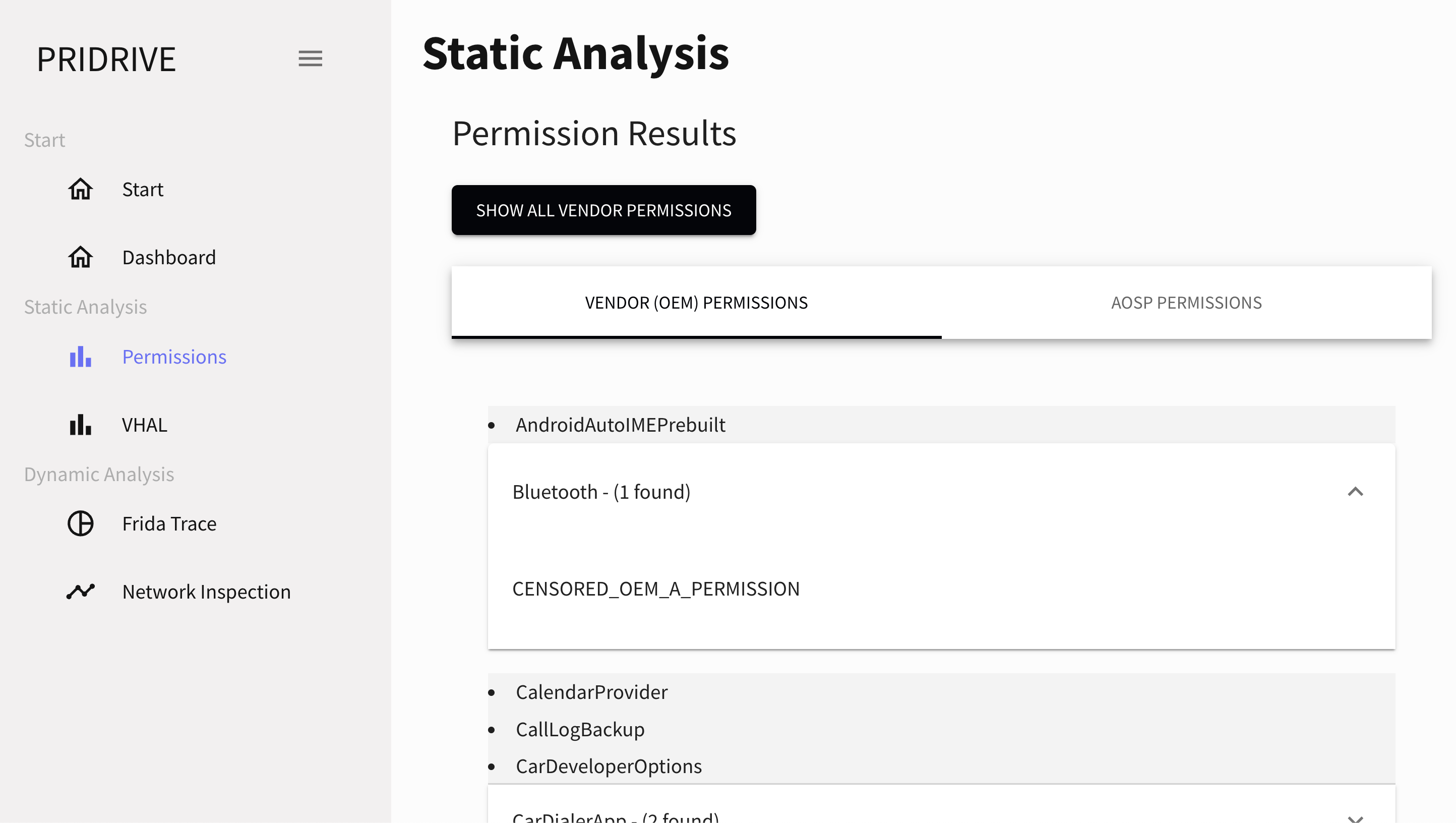}
    \caption{PriDrive Static Analysis Result Page}
    \label{fig:Pridrive_static}
\end{figure}

We have conducted a comprehensive data collection process involving three OEM emulator images. For each OEM, Our tool systematically parses and categorizes the current custom vehicle properties and permissions and those utilized by each APK.

% We have conducted a comprehensive data collection process involving four OEM emulator images. We developed a specialized Python script for each OEM to systematically parse and categorize the current custom vehicle properties and those utilized by each Android package. 

% \input{FiguresTex/tab_oem_permissions_vhal}

\textbf{Evaluation of Custom VHAL Properties.}
The interfaces between vehicle hardware and software functions are provided by VHAL properties. Each OEM has customized these features, despite this, we find patterns in the naming practices and identified similar VHAL attributes across various OEMs.

We extracted the top ten VHAL property IDs for three major OEMs which we have designated as \OEM{A}, \OEM{B} and \OEM{C} since we have not yet conducted a responsible disclosure. The following is an analysis of their data collection priorities and related privacy concerns.

% We extracted the top 10 VHAL property IDs for \OEM{A}, \OEM{B}, \OEM{C}, and \OEM{D}, aiming to analyze their data collection priorities and related privacy concerns.

\textbf{Similar Property Pairs.}
In order to identify similar VHAL properties across different OEMs, we employed the Jaccard similarity algorithm~\cite{jaccard}. This method helps us understand how extensive an OEM's property spectrum is compared to others, with a broader spectrum indicating a larger variety of properties that may be collected and shared.
% \mert{Why did we have to do it? Because classifying these properties is a subjective process. How many people did this?} 
To compare custom property names across properties from different OEMs, each custom property name was tokenized into a set of substrings separated by underscores. We established a minimal similarity score of 0.2 to filter out attributes that differ significantly from one another. Using this methodology, we were able to identify VHAL properties that, despite having different names, potentially perform related tasks or store data of similar types.

Subtracting similar property counts from the total count reveals a different set of properties compared to the other OEM. For between each OEM A and B, the Jaccard similarity $J(A, B)$, and different set of properties $max(0, A - J(A, B))$ can be found on Table \ref{tab:similar_properties}.

\label{subsec:appx_similar_prop_pairs}
\begin{table}[htbp]
\begin{tabular}{ccccc}
\hline
\textbf{Set 1} & \textbf{Set 2} & \makecell{\textbf{Similar Custom} \\ \textbf{Properties} \\ \textbf{J(A,B)}} & \makecell{\textbf{Different } \\ \textbf{Props of} \\ \textbf{Set 1}} & \makecell{\textbf{Different } \\ \textbf{Props of} \\ \textbf{Set 2}} \\
\hline
\OEM{A} & \OEM{C} & 437 & 316 & 0\\
\OEM{A} & \OEM{B}  & 105 & 648 & 330\\
\OEM{B} & \OEM{C} & 109 & 326 & 83\\
\hline
\end{tabular}
\caption{Summary of Similar Property Pairs}
\label{tab:similar_properties}
\end{table}

For \textbf{\OEM{A}}, we have observed that there are a large number of accesses to 'INFO-VIN' and various 'TPMS-STATUS.' Accessing the former property frequently indicates data is being collected and tracked with an identifier. Location data in the form of 'GPS-POSITION', 'GPS-TIME', and 'GPS-VELOCITY' are also retrieved from the subscription manager. 
%notifications, and bug report apps. 
Additionally, safety data such as 'DRIVER-SEAT-BELT-STATUS' and 'TPMS-STATUS-LF' (tire pressure, left front wheel) is accessed. Such data is described in their privacy policy as being shared with third parties, for example, financial organizations who offer financing for the purchase or lease of OEM vehicles or usage-based insurance providers. 'VEHICLE-MASS-ESTIMATE' and 'TRAILER-CONNECTION-STATUS' are also used in tracking vehicle payloads. Additionally, for EVs, several properties such as 'HIGH-VOLTAGE-BATTERY-PROPULSION-RANGE', 'TIME-OF-DAY-CHARGING-NEXT-PLANNED-TARGET-CHARGE-LEVEL', and 'LOCATION-BASED-CHARGING-CUSTOM-IZATION-CURRENT-SETTING-VALUE' are also accessed by background apps and may be used in tracking charging patterns or locations. The car also includes an Amazon Alexa app, which requests permissions to vehicle data such as 'CAR-ENGINE-DETAILED', 'CONTROL-CAR-CLIMATE', 'READ-CAR-STEERING', and more. The OEM also includes apps that request fine-grained location permissions. Finally, one recurring permission ('ACCESS-3RDPARTY') likely grants a 3rd party access to data.

\textbf{\OEM{B}}, 'CREATE-USER' and 'USER-IDENTIFICATION-ASSO-\\CIATION' appear to instantiate the user profile and tracking identifier used in other OEM apps such as CarSettings and CarSettings. On system boot, a call to the 'OEM-INITIAL-REGISTERED-VIN' occurs. Its prebuilt default Android Automotive Google Maps app request additional permissions such as 'CAR-ENERGY-PORTS', 'CAR-POWERTRAIN', and Google's 'AD-ID' in addition to the car make and car model information requested by the prior OEM's map implementation. It is likely that such information is used in Google's map UI and possibly their advertising services. The prebuilt Google Search app also requests permissions for a large amount of car properties, many that are not enabled by default. These properties pertain to the vehicles' tires, lighting, speed, ports, mileage, energy, exterior environment, and more. In our analysis, we observed that one of the properties accessed by Google Search was the 'HVAC-ELECTRIC-DEFROSTER-ON.'

\textbf{\OEM{C}}'s properties, such as 'DRIVER-SUPPORT-FUNCTION-ON' and 'CITY-SAFETY-WARNING-SENSITIVITY' also track potentially sensitive data. While this may be required for safety and support features, this could reveal driving patterns or habits, especially as these properties are accessed regularly. The OEM app permissions also request for checking payloads like with OEM A via 'TRAILER-PRESENT'. The OEM's Google Maps also requests 'CAR-IDENTIFICATION' and 'CAR-INFO' like other apps.

% \OEM{D}'s top properties show a focus on vehicle operating data, especially 'VEHICLE-SPEED-DISPLAY-UNITS.' Although gathering this kind of data can improve vehicle functionality and aid in performance monitoring, it can also raise privacy issues. It is imperative that Volvo establish a balance between collecting operational data and protecting the users' privacy when operating the car. 

% \subsubsection{Evaluation of Custom Permissions}

\subsection{Phase 2B: Dynamic Analysis}
\label{subsec:eval_dynamic_analysis}

\begin{figure*}[htbp]
    \centering
    \begin{subfigure}[b]{0.32\textwidth}
        \centering
        \includegraphics[width=\textwidth]{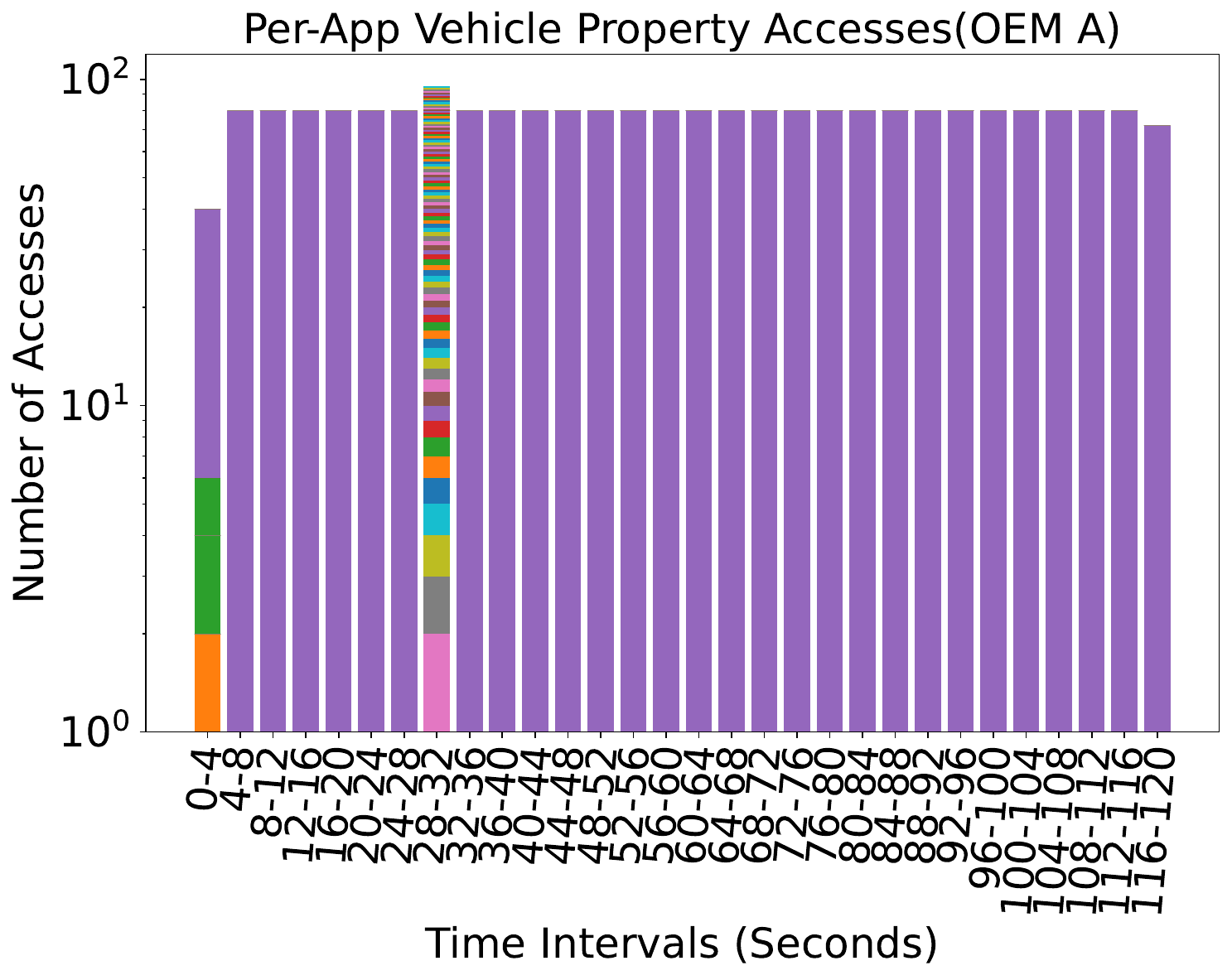}
        \caption{OEM A}
        \label{fig:gm_system_wide_prop}
    \end{subfigure}
    \hfill
    \begin{subfigure}[b]{0.32\textwidth}
        \centering
        \includegraphics[width=\textwidth]{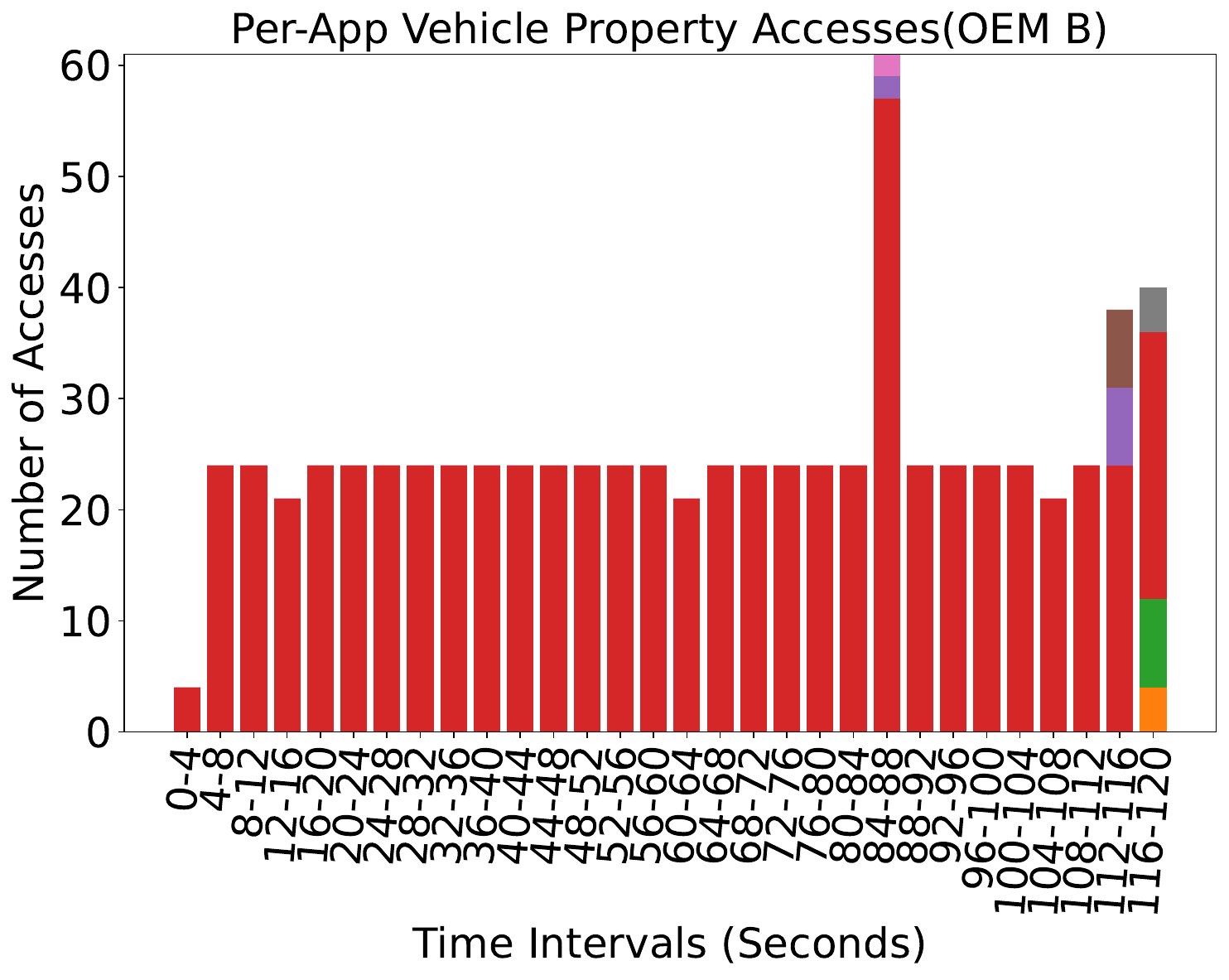}
        \caption{OEM B}
        \label{fig:honda_system_wide_prop}
    \end{subfigure}
    \hfill
    \begin{subfigure}[b]{0.32\textwidth}
        \centering
        \includegraphics[width=\textwidth]{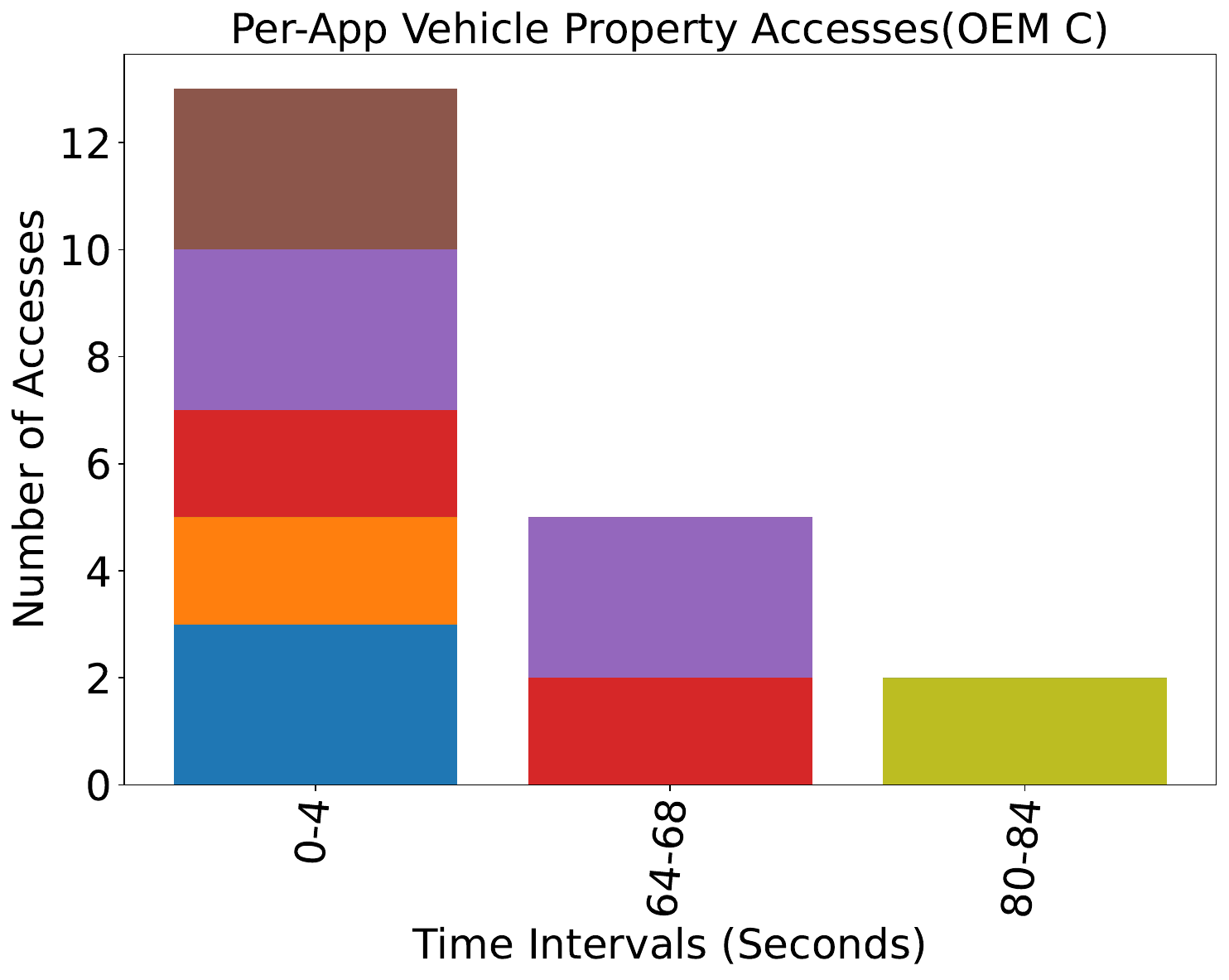}
        \caption{OEM C}
        \label{fig:pole_system_wide_prop}
    \end{subfigure}
    \caption{Frequency distribution of vehicle property accesses across all apps for different OEMs. Different colors represent different VHAL Property IDs. Purple on (a) represents Vehicle Speed Property ID, and red on (b) represents OEM-specific vehicle dynamics-related property that includes vehicle speed.}
\label{fig:system_wide_prop_access}
\end{figure*}

\begin{figure*}[htbp]
    \centering
    \begin{subfigure}[b]{0.32\textwidth}
        \centering
        \includegraphics[width=\textwidth]{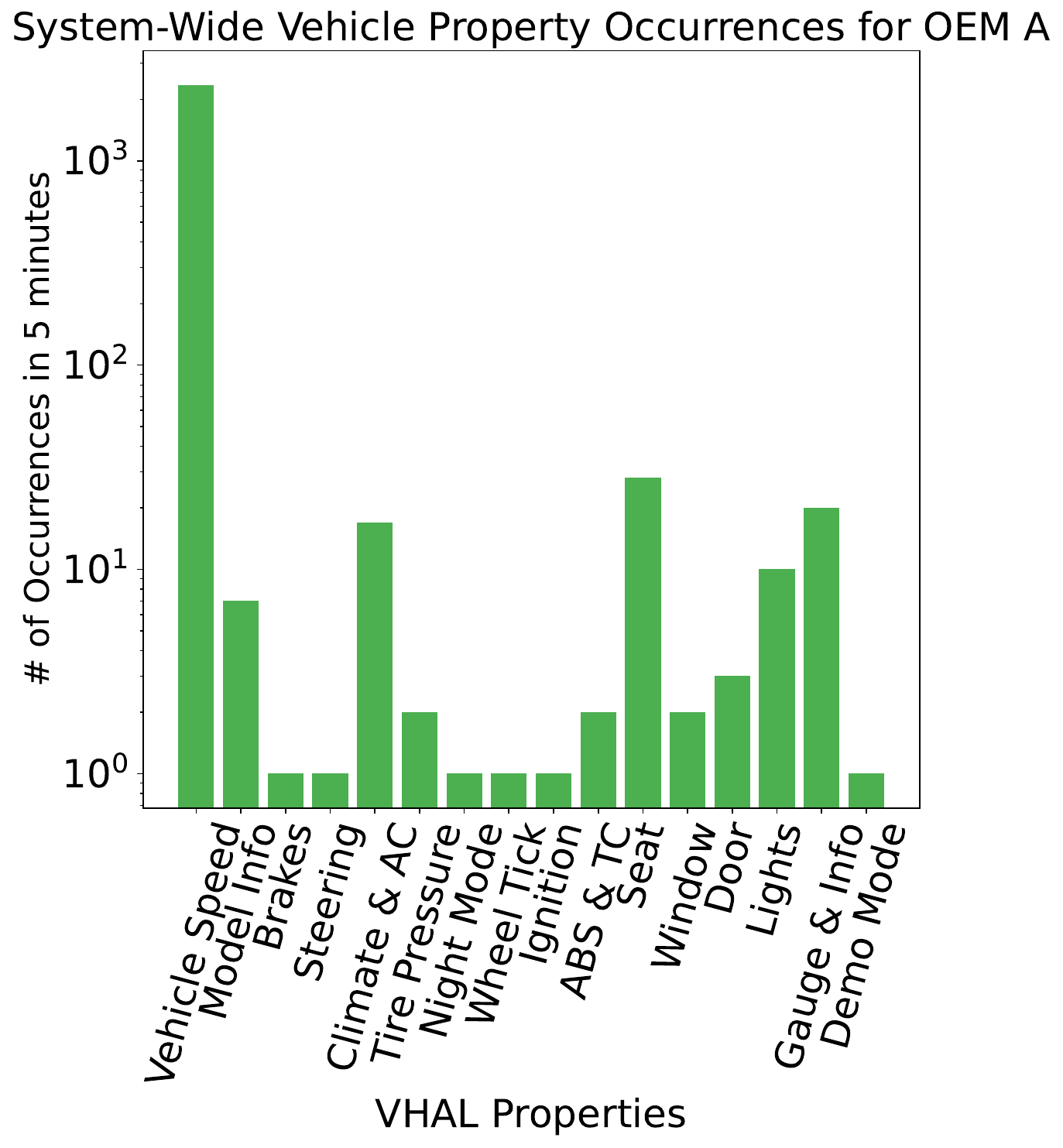}
        \caption{OEM A} 
        % \bulut{mention cumulative props}
        \label{fig:gm_system_wide_prop}
    \end{subfigure}
    \hfill
    \begin{subfigure}[b]{0.32\textwidth}
        \centering
        \includegraphics[width=\textwidth]{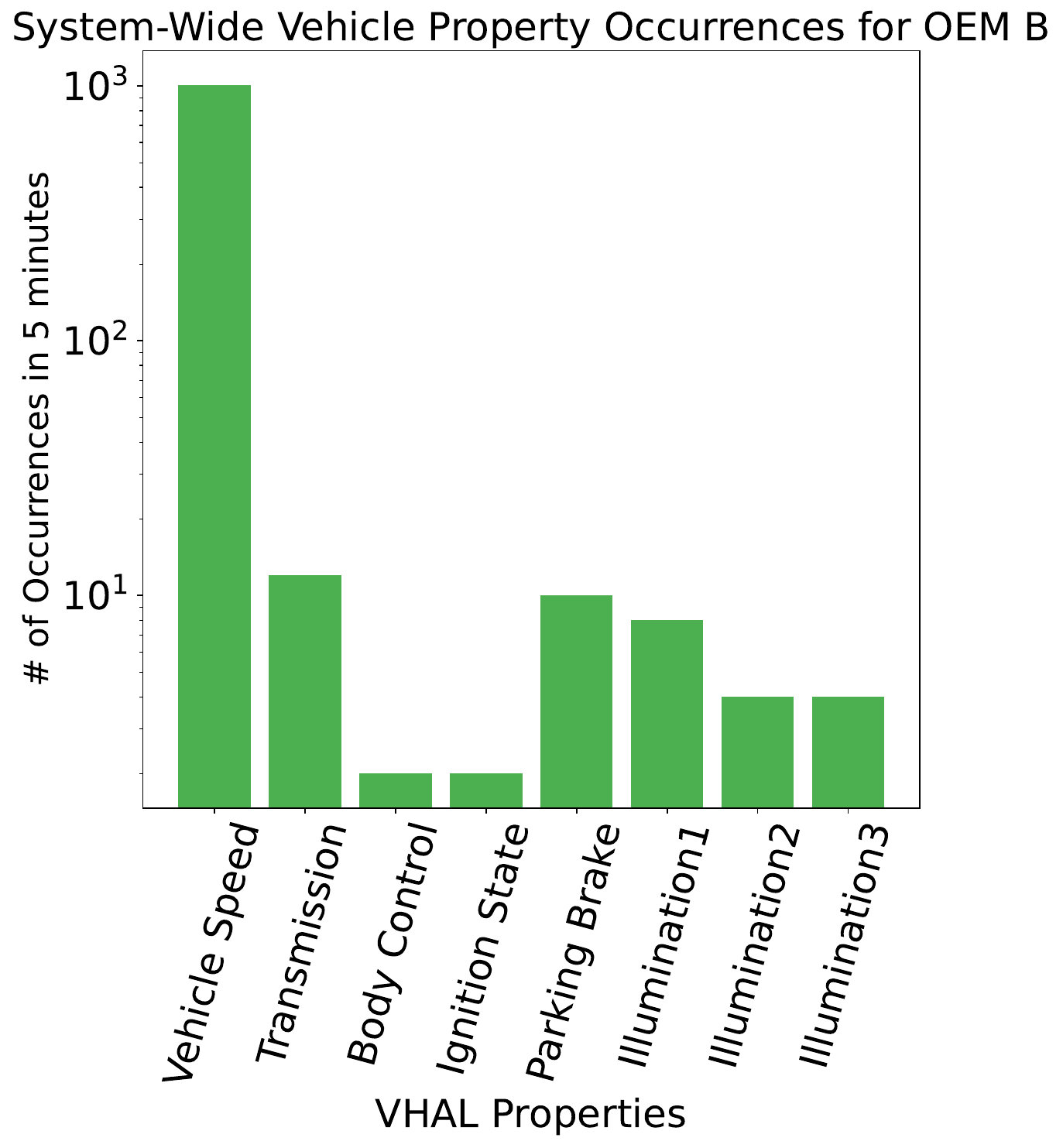}
        \caption{OEM B}
        \label{fig:honda_system_wide_prop}
    \end{subfigure}
    \hfill
    \begin{subfigure}[b]{0.32\textwidth}
        \centering
        \includegraphics[width=\textwidth]{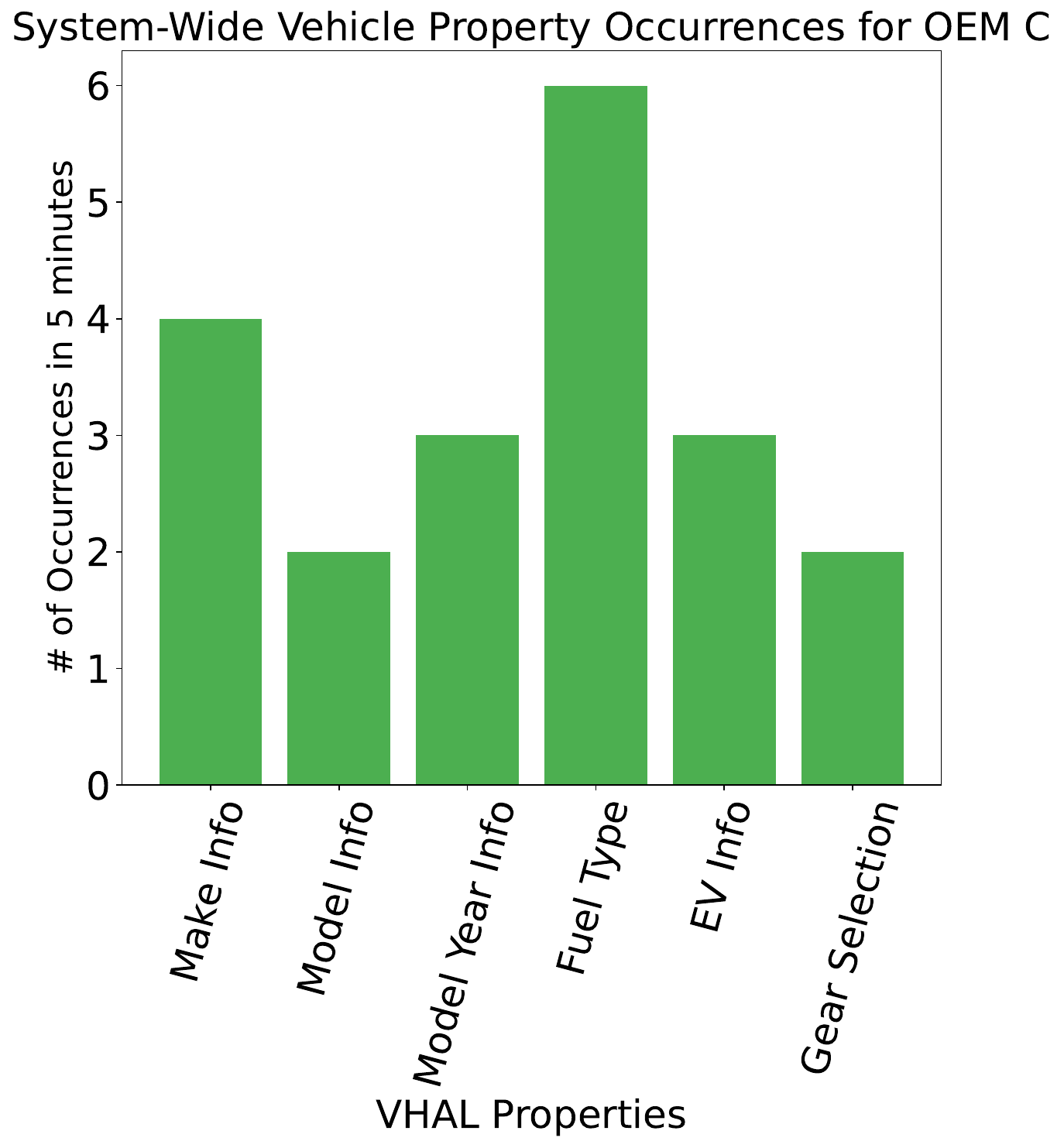}
        \caption{OEM C}
        \label{fig:pole_system_wide_prop}
    \end{subfigure}
    \caption{Frequency distribution of vehicle property accesses across all apps for different OEMs.}
    \label{fig:system_wide_prop_access}
\end{figure*}

Using the custom vendor property IDs discovered during the static analysis phase, we determined the frequency and the specific instances when a particular VHAL property was used within these five-minute timeframes. %And we have collected our data in JSON format. 
This helps us to systematically investigate the AAOS systems implemented by three major OEMs, OEM A, OEM B and OEM C, since we have not yet conducted a responsible disclosure. 
% \mert{You are already using this terminology in Phase 1 eval. Introduce the need for anonymization above, not here.} 
After an in-depth analysis, we were able to gather rich and diverse results, which could provide us with an understanding of the varied data collection practices employed by these different AAOS implementations.  

% We executed a detailed and comprehensive dynamic analysis by utilizing Frida trace outputs. This analysis was conducted over periods of five minutes for each application package. Using the custom vendor property IDs discovered during the static analysis phase, we determined both the frequency and the specific instances when a particular VHAL property was used within these five-minute timeframes." And we have collected our data in JSON format. This helps us to systematically investigate the Android Automotive systems implemented by four major Original Equipment Manufacturers (OEMs), which we have designated as OEM A, OEM B, OEM C, and OEM D, since we have not yet conducted a responsible disclosure. After an in-depth analysis, we were able to gather rich and diverse results, which could provide us with an understanding of the varied data collection practices employed by these four different Android automotive systems.  
%\textbf{OEM A.}
According to our collected traces, \textbf{\OEM{A}} has demonstrated a comprehensive approach to vehicle monitoring through the data collected in their Android Automotive system. This method includes multiple elements of vehicle functionality and driver engagement, such as:
\begin{itemize}
\itemsep0em 
  \item Information about the vehicle make, model and model year is collected in the Google Maps application. These properties could be aggregated to easily identify the exact vehicle.
  \item \OEM{A} monitors their vehicles in real-time, based on the data from their AAOS, especially the vehicle speed, which is captured approximately 20 times per second. The speed property could be used to infer other data, such as stops, turns, driving habits, or risky driving behaviors.
  \item The status of the parking brake and various HVAC (Heating, Ventilation, and Air Conditioning) settings, including maximum AC, fan direction, fan speed, and seat temperature, are monitored, which are comfort attributes.
  \item Data on turn signal state, tire pressure, and even the state of fog lights are recorded. This highlights \OEM{A}'s concentration on detailed driving dynamics and vehicle status.
  \item The system also captures more specialized data, such as the headlight and hazard light states, seat adjustments, and the status of electric vehicle (EV) charge ports.
  \item Vehicle state information is collected regularly upon startup via a batched data dump.
\end{itemize}
\OEM{A} monitors multiple elements of vehicle operation through considerable property accesses, which include a substantial number of unique vendor-specific property IDs. 

%\subsubsection{OEM B}

The dynamic analysis of \textbf{\OEM{B}}'s AAOS monitored a range of vehicle-specific functionalities. The VHAL properties we run into include:
\begin{itemize}
\itemsep0em 
  \item Changes in illumination settings signify that the system is modifying properties to visibility and lighting inside the car.
  \item Memory features for daytime lighting preferences demonstrate the ability to retrieve user-specific settings.
  \item Changes to the lighting settings that correspond to the vehicle's lighting system are being adjusted in real time.
  \item Parking brake status is a sign for determining where the vehicle is parked.
  \item Ignition and transmission state offer perceptions about the usage and operational state.
  \item Body control functions, which may indicate accessing various body-related features.
  \item Engine and vehicle speed, occurred with high frequency (approximately five times per second), highlighting a significant focus on real-time speed monitoring.
\end{itemize}
The examination of \OEM{B}'s automotive system reveals attention to specific vehicle characteristics such as high-frequency engine and vehicle speed monitoring.

%\subsubsection{OEM C}

In the analysis of \textbf{\OEM{C}}'s Android Automotive system, the data captured includes VHAL property IDs corresponding to:
\begin{itemize}
\itemsep0em 
  \item Gear selection status, which may provide insights about driving patterns.
  \item Types of EV charger connectors used may be relevant for understanding interactions with charging infrastructures.
  \item Details about fuel types that reveal the energy source of the car.
  \item Data about the vehicle make, model year and particular model.
\end{itemize}
\OEM{C} mostly accesses VHAL properties related to energy and vehicle information. Collecting extensive amounts of these types of data might raise privacy concerns, particularly when linked to user profiles or driving habits.

OEM A, by far, collects the most vehicle data with properties relating to vehicle systems, user preferences, driver assistance data, seatbelt information, and other safety and performance data. While the other OEMs collect fewer properties, they still each track a VIN and retrieve speed data, make/model data, and more. Furthermore, OEM A data are aggregated into categories in \ref{fig:system_wide_prop_access}. This is because OEM A's properties include details of some parameters as different properties, which would be difficult to represent in the plots.

% \subsubsection{OEM D}

% We have obtained a dynamic analysis trail from \OEM{D}'s data, but we were unable to locate any instances of access to a vendor property ID. There could be a number of reasons for this. This might indicate that, in comparison to other OEMs, the vehicle's system either doesn't access these properties within the five-minute interval examined or does so less frequently.  Another explanation for the ineffectiveness of the data collection technique on \OEM{D}'s system could be technical constraints or the existence of security mechanisms that identify and prevent analytical tools. 

\subsection{Phase 2C: Network Traffic Inspection}

\textit{Wireshark} and \textit{mitmproxy} were utilized for our analysis. \textit{Mitmproxy} revealed the HTTPS traffic associated with AAOS, while \textit{Wireshark} provided insights into the details of the network transmission. In this part, we have discussed the results of using these tools.

\begin{figure*}[ht]
    \centering
    \includegraphics[width=\textwidth]{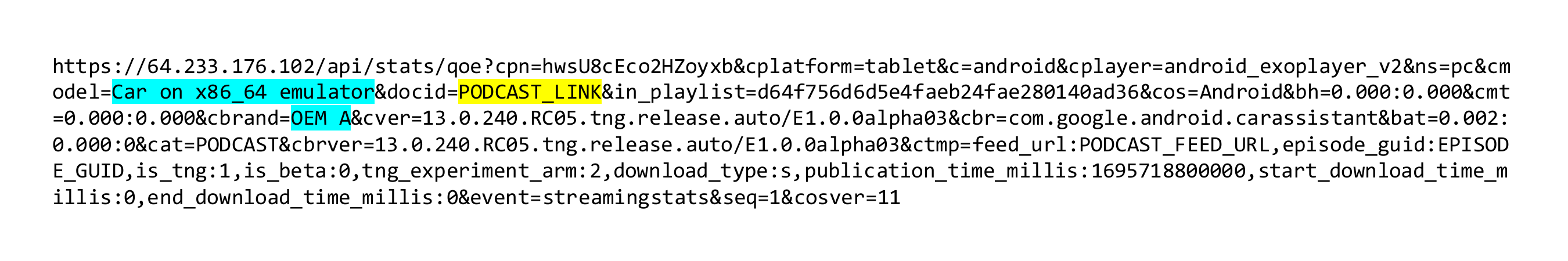}
    \vspace{-12mm}
    \caption{A Get Request sent to Google Servers}
    \label{fig:get_request}
\end{figure*}

We examined HTTPS traffic that was decrypted using an MITM approach, discovering that much of the data is encoded. Additionally, some of the data is in Protocol Buffers (protobuf) format. Using the protoc executable, we can convert the encoded data into a readable format. However, since there is no clear understanding of the key-value pairs in protobuf, we were unable to extract meaningful information. Conversely, we identified human-readable strings in the traffic that do not pose any privacy concerns. In addition, we have formed graphs about the total payload size in 5 minutes of recording for each OEM. 

\begin{figure*}[!bp]
    \centering
    \begin{subfigure}[b]{0.69\columnwidth}
        \centering
        \includegraphics[width=\textwidth]{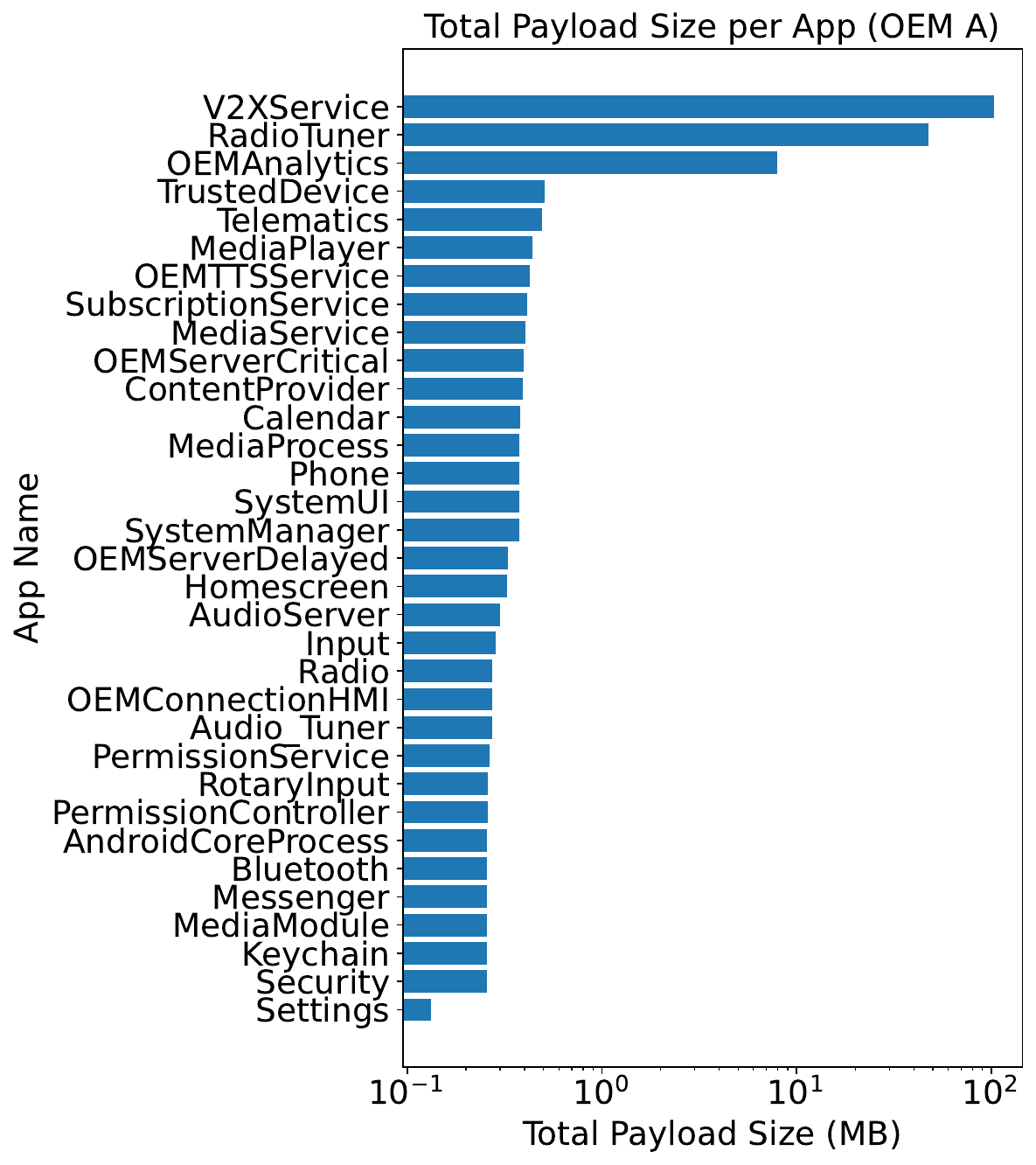}
        \caption{Total HTTPS Payload Size per App in 5 minutes of recording (OEM A)}
        \label{fig:gm_payload_size}
    \end{subfigure}
    \hfill
    \begin{subfigure}[b]{0.69\columnwidth}
        \centering
        \includegraphics[width=\textwidth]{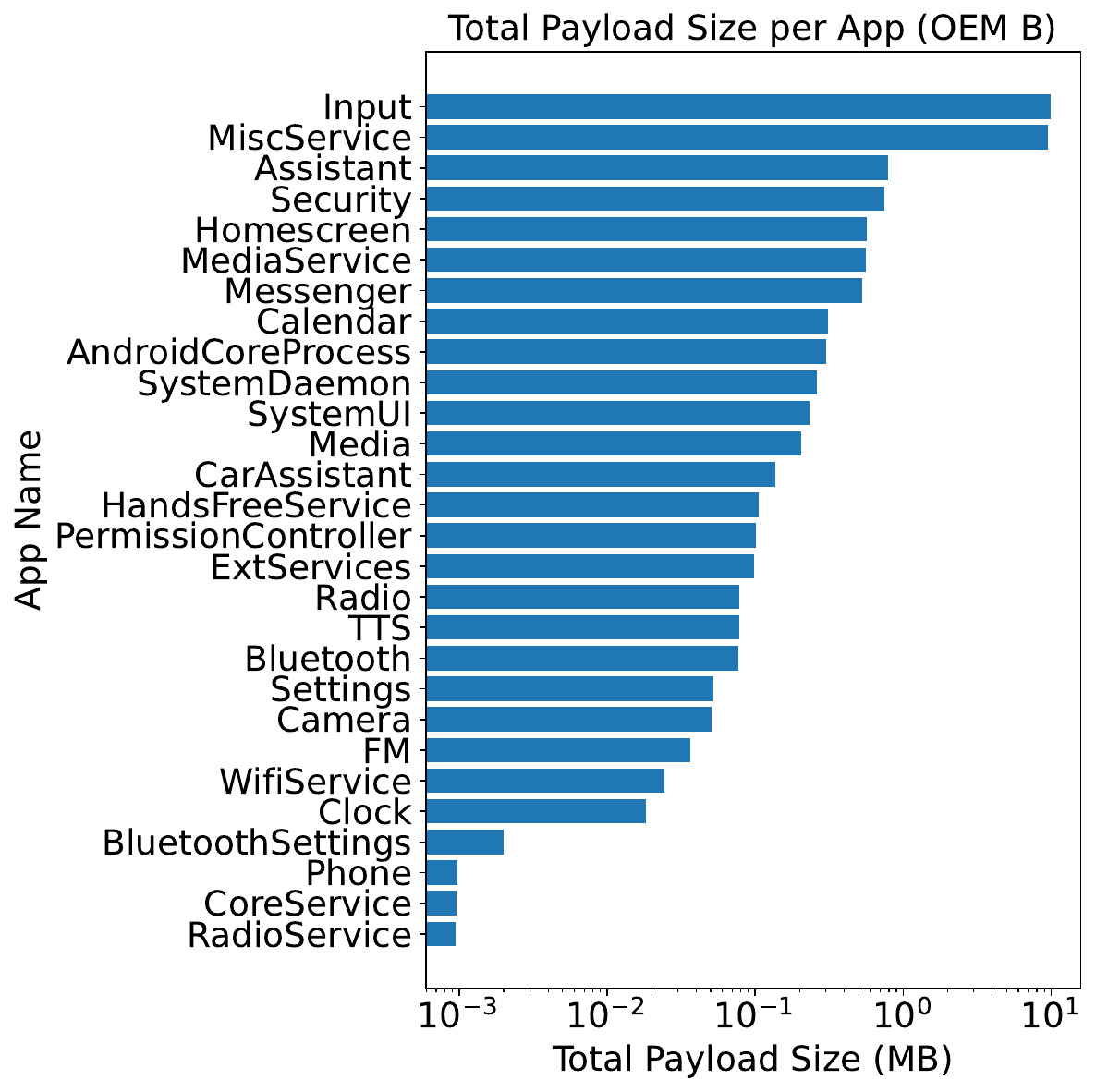}
        \caption{Total HTTPS Payload Size per App in 5 minutes of recording (OEM B)}
        \label{fig:honda_payload_size}
    \end{subfigure}
    \hfill
    \begin{subfigure}[b]{0.69\columnwidth}
        \centering
        \includegraphics[width=\textwidth]{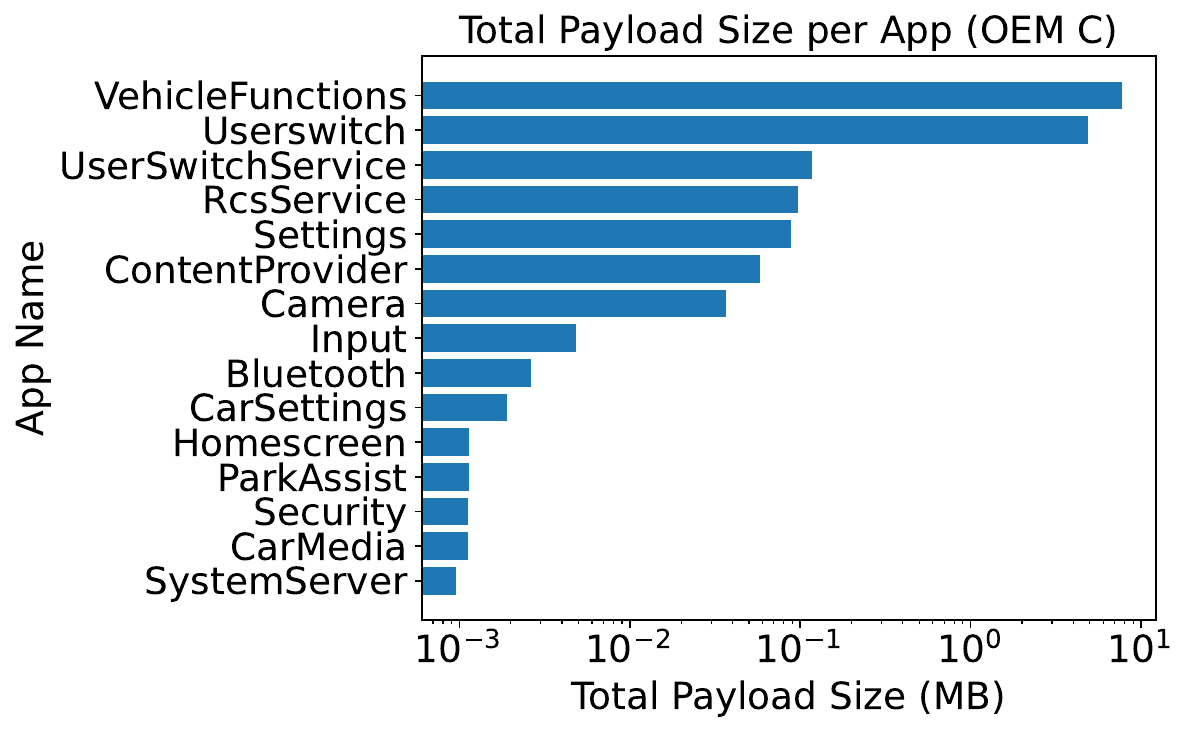}
        \caption{Total HTTPS Payload Size per App in 5 minutes of recording (OEM C)}
        \label{fig:pole_payload_size}
    \end{subfigure}
        \caption{Per-app network traffic requests over time for different OEMs.}
    \label{fig:payload_size_per_app}
\end{figure*}

The apps that send the most data over HTTPS in each OEM include analytics and services apps. These apps use significantly more network traffic compared to prebuilt Android apps. OEM A's V2XService, RadioTuner, and OEMAnalytics apps all send tens to hundreds of MB of data to servers. Upon conducting reverse IP lookups, a majority of IPs are Google datacenters in California, Ireland, Singapore, and other locations.

Furthermore, our findings revealed that streaming statistics sent to Google Servers as GET requests include the car model and the media content link, as shown in Figure \ref{fig:get_request} (blue highlights are device information, and yellow highlight is the content link). This data transmission occurs during media playback. Thus, Google can associate specific media content with particular vehicle models. Assuming that Google also has the user's personal identification data, this precise linkage between car models, media consumption, and personal identification patterns may pose a significant privacy risk. PODCAST\_LINK is censored to avoid copyright issues.

\subsection{Phase 2D: Privacy Policy Evaluation}
\label{sec:privacy}

\textbf{OEM Privacy Policies}
A privacy policy is said to be inconsistent with a company's data collection practices if the data collection of a vehicle property is not properly disclosed in the privacy policy. For example, if an OEM claims to collect personal data, vehicle information, and vehicle performance data, but does \textit{not} claim to collect and use driver safety data, we should expect a vehicle property such as ``DRIVER\_SEAT\_BELT\_STATUS'' or ``LANE\_DETECTION\_WARNING'' to not be collected by the OEM. This is considered an \textit{omission} of data practices and indicates an inconsistency between data collection and privacy disclosure.

We collected linked privacy policy pages and PDFs on each OEM's website and automatically parsed each of them, generating data flows for each sentence. The privacy policy documents included the following: General Privacy Policy, California Statements, US Consumer Privacy, US Connected Services, Linked 3rd Party Policies, OEM App Privacy Policies, Speech Messaging Policy, and Real Time Traffic Info Privacy Policy.

\begin{table}[]
\centering
\begin{tabular}{|l|l|l|l|}
\hline
\rowcolor[gray]{0.8}
  & \textbf{Catg. Disclosed} & \textbf{Prop. Disclosed} & \textbf{Purposes} \\ \hline
OEM A & 15/16 (93.75\%) & 735/845 (86.98\%) & 23 \\ \hline
OEM B & 13/15 (86.67\%) & 41/67 (61.19\%) & 23 \\ \hline
OEM C & 4/7 (57.14\%) & 8/33 (24.24\%) & 19 \\ \hline 
\end{tabular} 

\caption{Privacy policy consistency rates with VHAL properties and high-level data categories.}
\label{tab:policycompliance}
\end{table}

\textbf{OEM A. }
For OEM A, the privacy policies were not detected to have mentioned the data category of vehicle climate and comfort information, such as seat settings and HVAC settings. This covers 110 vehicle properties, or 13.02\% of properties found in the static analysis.
Their privacy policies discuss a number of issues regarding data collection and usage practices. For example, they discuss using data to facilitate financial organizations and for financing with purchasing vehicles or enrolling in usage-based insurance providers. For vehicle usage information, they collect both driving behavior, location, and multimedia data for safety and performance evaluations, as well as tracking user location for providing location based services. In their policy, they indicate that they do not sell collected voice biometric data \textit{without consent}, though the requirements for consent may be included in the terms of purchasing a vehicle. They also share information with law enforcement and roadside assistance providers to ensure safety and provide services for assisting in recovering stolen vehicles. OEM A uses the onboard cameras and sensor data to ``enhance vehicle functionality''.

\textbf{OEM B.}
Our tool also did not detect any disclosure of using vehicle climate and comfort information in OEM B, as well as vehicle lighting info, covering 26 (38.81\%) properties. 

OEM B collects data for advertising and marketing of model-specific hybrid accessories for product recommendations and sales. Additionally, OEM B collects data upon consent to facilitate services that rely on geolocation information (such as traffic, map, navigation, driver behavior, and geo-fencing). They provide fuel-saving services and also congestion calculation (using vehicle position, speed, and VIN). They also collect voice recognition data for product improvement.

\textbf{OEM C.}
OEM C was found to not disclose the usage of driving assistance information, vehicle climate and comfort information, and vehicle audio and video information, which covers 25 (75.76\%) of the accessed vehicle properties.

They engage in lookalike advertising, target marketing channels, and provide personalized content and marketing leads. The OEM engages in data collection for user profiling and targeted advertising on Facebook. OEM C discloses the usage of identifiers, commercial information, and geolocation data to service providers and insurance companies for their driver feedback program. Like OEM A, they collect VINs and location data for roadside assistance, and voice commands and voice biometric data. They collect general data to assist law enforcement and data from onboard cameras to develop products. Unlike, OEM A and B, they collect location information and electricity consumption data for trip details.

%\subsection{Phase 2E: 3rd Party Apps}

\section{Evaluation of Third-Party Apps}
\label{sec:eval_third_party}

Our methods can be applied to Android applications developed by third parties as well. We have compiled a selection of the most widely used third-party applications from the Google Play Automotive Store. The same process is followed as discussed before. As a result, we got the table of each 3rd party app's access to different VHAL property IDs. Although some counts are relatively high or low compared to others, the 46 distinct counts of unique VHAL parameters may be due to governmental regulations or privacy concerns, which are observed on 12 different applications (see Table \ref{tab:3rdparty_stat}).

\begin{table}[h]
\centering
\begin{tabular}{ll}
\toprule
\textbf{Application} & \makecell{\textbf{Unique Property} \\ \textbf{Occurrence}} \\ \midrule
com.coulombtech & 128 \\ \hline
com.waze & 47 \\ \hline
com.iternio.abrpapp & 47 \\ \hline
com.outdooractive.Outdooractive & 46 \\ \hline
energy.octopus.electricjuice.android & 46 \\ \hline
com.telenav.app.android.scout\_us & 46 \\ \hline
com.parkwhiz.driverApp & 46 \\ \hline
com.kikkoapps.parkingcheapsg.free & 46 \\ \hline
net.vonforst.evmap & 46 \\ \hline
com.stuzo.texaco & 46 \\ \hline
com.xatori.Plugshare & 46 \\ \hline
com.stuzo.chevron & 46 \\ \hline
com.acmeaom.android.myradar & 46 \\ \hline
com.trydriver.driver.automotive & 46 \\ \hline
com.crosschasm.evchargerlocator & 46 \\ \hline
twc.car.tmpl.go & 45 \\ \hline
no.tv2.sumo & 29 \\ \hline
org.pbskids.video & 12 \\ \hline
com.spothero.spothero & 9 \\ \hline
com.spotify.music & 3 \\ \hline
com.amazon.avod.thirdpartyclient & 2 \\ \hline
com.clearchannel.iheartradio.controller & 2 \\ \bottomrule \\ 
\end{tabular}
\caption{ Occurrences of Unique VHAL Properties Across Third-Party Applications from Google Play Store for AAOS}
\label{tab:3rdparty_stat}
\end{table}
%-------------------------------------------------------------------------------
\section{Discussion}
\label{sec:discussion}
%-------------------------------------------------------------------------------

%We propose a system designed to detect privacy violations within 
%Android Automotive OS infotainment systems. These systems, 
%developed by manufacturers such as Harman \cite{harman} and 
%Snapp Automotive at the request of OEMs, are different from 
%those produced by the automotive manufacturers themselves. 
%By integrating our system, OEMs can effectively assess and 
%monitor the privacy conditions of the software and systems 
%they deploy.

\subsection{Recommendations}
\label{sec:recommendations}

Based on our research, we think OEMs could take inspiration from other platforms, like the web, to give their customers a more safe and privacy-friendly experience. In particular, we suggest the following actions in light of our findings:
\begin{itemize}
    \item OEMs should give their customers the option to opt out of data collection by their services while still allowing the product (i.e., infotainment system) to be functional. Since the privacy policies are often presented at the dealership when buying a new vehicle, the dealer could be incentivized to discuss the OEM's data collection practices with the customer before enabling AAOS.
    \item Since OEMs already offer a \textit{walled} Play Store with third-party apps of their own choosing, the vetting process of these apps can be further enhanced by adding privacy considerations. An app with privacy policy inconsistencies above a certain threshold (which can be as small as zero) can be returned for rework or declined by the OEM.
    \item The lack of car-specific legislation, as well as their geographic fragmentation, can be attributed to the lackluster implementation and consideration of privacy in the automotive context. However, increasing privacy regulation frameworks such as the European Union's General Data Protection Regulation (GDPR) will affect any OEM that sells vehicles in the important European market. OEMs are considered data controllers according to GDPR because they have control over the data that third-parties can access and use. This means that they have more compliance requirements, and they are directly in charge of putting in place the necessary  countermeasures to avoid costly litigation.
\end{itemize}

\subsection{Limitations and Future Work}
\label{sec:futurework}

We have collated data through static analysis, dynamic analysis, 
and network traffic inspection. We plan to open-source \name that is capable of autonomously initiating 
emulators, rooting them, and conducting static analyses. 
The process would be systematically designed to analyze APK 
packages present within each emulator. The system would then
seamlessly transition to perform dynamic analysis and network 
traffic inspection. In future, we may consider fully 
automating the interaction of screen components and simulating 
complex vehicle states, enabling a more in-depth and realistic 
measurement study. Another challenge in this research involves 
the decoding of protocol buffers. Addressing this using 
statistical analysis approaches may be able to allow us to 
confirm exactly what data is sent to Google and OEM servers.

We plan to interconnect these various components into a 
public-facing application for OEMs and Android Automotive to 
automatically check and ensure data compliance. 
Ideally, the existing AAOS permissions system could be 
overhauled to include stricter permissions requirements and 
explicit opt-in consent for certain data types, rather than 
requiring all data types to be collected for the vehicle 
services to function. Our modified permissions system would also 
provide a brief privacy summary related to data usage purposes 
and the sensors or sources of data collection with ratings for 
sensitive vehicle property attributes.

%-----------------------------------------------------------------
\section{Conclusion}
\label{sec:conclusion}
%-----------------------------------------------------------------

In this study, we collected an extensive dataset from three distinct emulators, focusing on the custom Android permissions of OEMs, their unique VHAL property IDs, and the frequency of their usage. Additionally, we have meticulously gathered data regarding the network behavior and runtime logs of every APK application package involved in our study. Furthermore, it is important to note that the collection of this data does not inherently signal privacy concerns. However, our efforts represent significant progress in thoroughly analyzing AAOS's privacy and security aspects. This study lays the groundwork for future investigations into potential privacy issues and security vulnerabilities within this domain, thereby contributing to enhancing user privacy and security in the evolving field of automotive technology.

% In this study, we have successfully collected an extensive dataset from four distinct emulators, focusing on the custom Android permissions of OEMs, their unique VHAL property IDs, and the frequency of their usage. Additionally, we have meticulously gathered data regarding the network behavior and runtime logs of every APK application package involved in our study. Furthermore, it is important to note that the collection of this data does not inherently signal privacy concerns. However, our efforts represent significant progress in thoroughly analyzing AAOS's privacy and security aspects. This study lays the groundwork for future investigations into potential privacy issues and security vulnerabilities within this domain, thereby contributing to enhancing user privacy and security in the evolving field of automotive technology.

\newpage

%\begin{acks}
%To Robert, for the bagels and explaining CMYK and color spaces.
%\end{acks}

\bibliographystyle{ACM-Reference-Format}
\bibliography{refs}

\appendix
\section{Vehicle Data Categories}
%\label{sec:appdx}

%\subsection{Vehicle Data Categories}
\label{subsec:appdx_data_categories}

\textbf{User Preferences and Notifications Settings.}
This category deals with customization and user-specific settings in the car, both of which are essential for improving the user experience. However, it also raises privacy issues because a user's personal choices might expose a lot about their lifestyle and habits.

\textbf{Driving Assistance and Mode Security.}
This area focuses on the vehicle's security features and driving assistance systems, like lane-keeping assistance and adaptive cruise control. These features are essential for lowering accident rates and improving driving safety. However, there is a privacy issue because driving behaviors and aid usage data can be sensitive and show drivers' abilities and habits.

\textbf{Energy and Maintenance.}
This category includes data related to vehicle energy consumption, battery health, and maintenance needs, which are vital in automotive engineering for predictive maintenance and energy efficiency, essential to sustainable automotive technology. However, a privacy concern is associated with this as it can provide insights into the vehicle's usage patterns and mechanical state, which might be considered private.

\textbf{Lighting.}
Vehicle lighting configurations have functions beyond aesthetics; they greatly enhance comfort and safety. On the other hand, there is also a privacy aspect to consider: while seemingly unimportant, lighting choices may be leveraged to deduce a user's location or time of trip.

\textbf{Diagnostic and Monitoring.}
Vehicle functionality and reliability are primarily dependent on ongoing diagnostics and monitoring. In terms of privacy, diagnostic data might provide comprehensive details about the state and use of the vehicle. Information on how, when, and where the vehicle is used could be in this potentially sensitive data, which could result in privacy violations if improperly handled and safeguarded.

\textbf{Climate and Comfort.}
The settings for the climate control system greatly influence the comfort of a vehicle's passengers. Yet, these options also have a privacy component; for example, temperature preferences may be used to infer personal information about the occupants, such as their preferred level of comfort or even their health.

\end{document}